\documentclass[usenatbib]{mn2e}
\usepackage{amssymb}
\usepackage{times}
\usepackage{epsfig}

\newcommand{\rmd}{{\rm d}}
\newcommand{\msun}{{\rm M}_{\sun}}
\newcommand{\xte}{{\it RXTE}}
\newcommand{\sax}{J1808}
\newcommand{\igr}{IGR~J00291+5934}
\newcommand{\xmm}{{\it XMM-Newton}}
\newcommand{\xspec}{{\sc XSPEC}}

\newcommand{\kte}{T_\mathrm{e}}
\newcommand{\ktseed}{T_\mathrm{seed}}
\newcommand{\ktbb}{T_\mathrm{bb}}
\newcommand{\fcol}{f_\mathrm{col}}
\newcommand{\rg}{r_\mathrm{s}}
\newcommand{\rstar}{R_\ast}
\newcommand{\rapp}{R_\infty}
\newcommand{\rco}{R_\mathrm{cor}}
\newcommand{\ka}{k_\mathrm{A}}
\newcommand{\rin}{R_\mathrm{in}}
\newcommand{\sigmat}{\sigma_\mathrm{T}}
\newcommand{\taut}{\tau_\mathrm{T}}
\newcommand{\mpro}{m_\mathrm{p}}
\newcommand{\refl}{{\Re}}
\newcommand{\ani}{h}
\newcommand{\shard}{\Sigma_\mathrm{hard}}
\newcommand{\ssoft}{\Sigma_\mathrm{soft}}
\newcommand{\beq}{\begin{eqnarray}}
\newcommand{\eeq}{\end{eqnarray}}
\newcommand{\be}{\begin{equation}}
\newcommand{\ee}{\end{equation}}

\begin{document}
\title[Accreting millisecond pulsar SAX J1808.4$-$3658]{Accreting millisecond pulsar SAX J1808.4$-$3658 during its 2002 outburst:
evidence for a receding disc}

\author[A. Ibragimov and J. Poutanen]
{Askar Ibragimov$^{1,2}$\thanks{E-mail:
askar.ibragimov@oulu.fi, juri.poutanen@oulu.fi }
and Juri Poutanen$^1$\footnotemark[1]  \\
$^1$Astronomy Division, Department of Physical Sciences, PO Box 3000, FIN-90014 University of Oulu, Finland\\
$^2$Kazan State University, Astronomy Department, Kremlyovskaya 18, 420008 Kazan, Russia }

\pagerange{\pageref{firstpage}--\pageref{lastpage}}
\pubyear{2009}
\date{Accepted 2009 July 31. Received 2009 July 30; in original form 2008 December 29}

\maketitle
\label{firstpage}
\begin{abstract}
An outburst of  the accreting X-ray millisecond pulsar SAX J1808.4$-$3658 in October--November 2002 was followed by the {\it Rossi X-ray Timing Explorer} for more than a month. A detailed analysis of this unprecedented  data set is  presented. For the first time, we demonstrate how the area covered by the hotspot at the neutron star surface is decreasing  in the course of the outburst together with the reflection amplitude. These trends are in agreement with the natural scenario, where the disc inner edge is receding from the neutron star as the mass accretion rate drops. These findings are further supported by the variations of the pulse profiles, which clearly show the presence of the secondary maximum at the late stages of the outburst after October 29.
This fact can be interpreted as the disc receding sufficiently far from the neutron star to open the view of the lower magnetic pole. 
In that case, the disc inner radius can be estimated.
Assuming that disc is truncated  at the Alfv\'en radius, we constrain  the stellar magnetic moment to $\mu=(9\pm5)\times10^{25}$ G cm$^3$, which corresponds to the surface field of about $10^8$ G.
 On the other hand, using the magnetic moment recently obtained from the observed pulsar spin-down rate we show that
the disc edge has to be within factor of two of the  Alfv\'en radius, putting interesting constraints on the models of the disc-magnetosphere interaction. 
We also demonstrate that the sharp changes in the phase of the fundamental are intimately related to the variations of the pulse profile, which we associate with the varying obscuration of the antipodal spot. Using the phase-resolved spectra, we further argue that  the strong dependence of the  pulse profiles  on photon energy and the observed soft time lags result from the different phase dependence of the  normalizations of the  two spectral components, the blackbody and the Comptonized tail,  being consistent with the model, where these components  have significantly different  angular emission patterns. The  pulse profile amplitude allows us to estimate the colatitude of the hotspot centroid to be $\sim4\degr$--$10\degr$.
\end{abstract}


\section{Introduction}

SAX J1808.4$-$3658 (hereafter \sax) is the first detected accretion-powered millisecond pulsar (AMSP) \citep{itz98,WvdK98}. \sax\ experiences  outbursts lasting a few weeks roughly once in two years, during which the coherent $\sim$401 Hz pulsations are observable. The neutron star accretes matter from a $\sim 0.05 M_\odot$ brown dwarf \citep{bc01} at  a 2 hr orbit \citep{CM98}. The magnetic field strength is estimated to be  about $10^8$--$10^9$ G \citep{PC99}. A recent analysis of the pulsar spin-down between the outbursts gives a narrower range of $B\sim$(0.4--1.5)$\times10^8$ G \citep[][ hereafter H08]{HPC08}. Among eight AMSP discovered since 1998 showing coherent pulsations over extended intervals during their outbursts (see \citealt{W06,p06} for reviews and \citealt{KMD07}), \sax\ is the best studied thanks to its five outbursts and their good coverage with the {\it Rossi X-ray Timing Explorer (\xte)}.

 \begin{figure*}
\centerline{\epsfig{file= 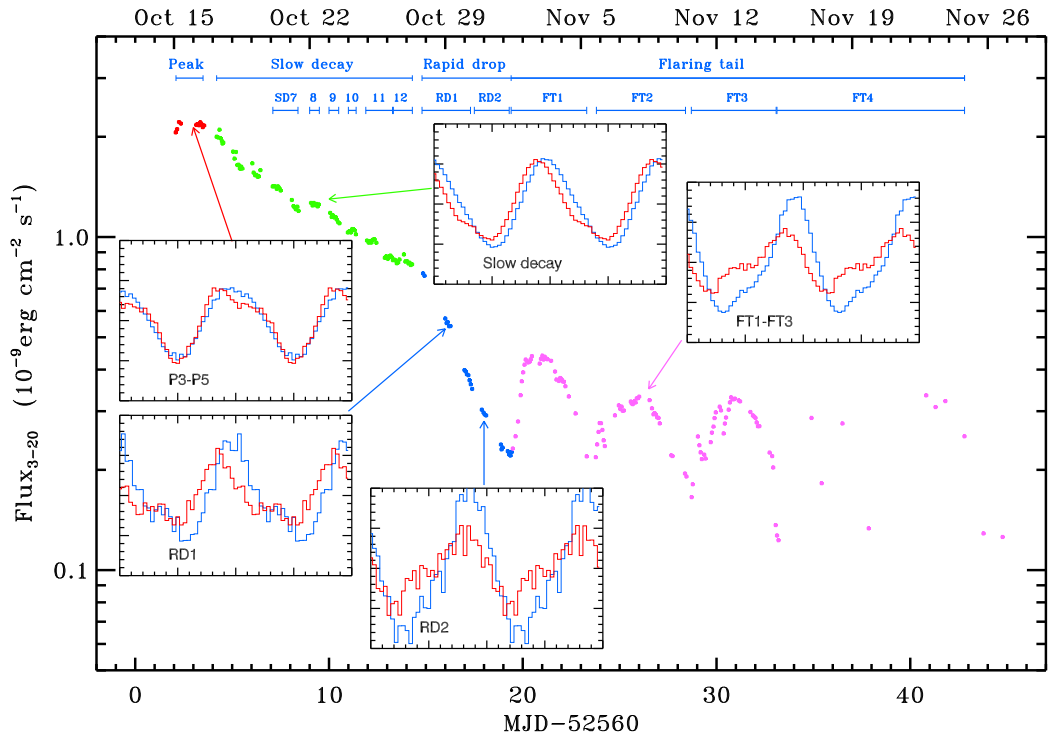,width=14cm}}
\caption{The light curve of  SAX J1808.4$-$3658 during the 2002 outburst.
The flux is computed in the 3--20 keV energy band.
We divide the outburst into 4 stages: peak (P), slow decay (SD), rapid drop (RD) and flaring tail (FT), which are coloured in  red, green, blue and magenta, respectively. Stripes indicate different outburst stages, see Table \ref{t:grouping}. The inserts show the pulse profiles at different times in the  2--3.7  and 10--24 keV energy bands (blue and red histograms, respectively).
}
\label{fig:overview}
\end{figure*}

The analysis of the broad-band spectra of \sax\ reveals the presence of at least two major components: soft, blackbody-like emission below 7 keV and a powerlaw tail with the cutoff at $\sim$100 keV (\citealt*{GDB02}; \citealt{PG03}, hereafter PG03). Both components are pulsating at the pulsar frequency and therefore are associated with the impact of the accretion stream to the neutron star surface. The hard powerlaw is most probably produced in the accretion shock, while the blackbody might be the heated neutron star surface underneath and around this shock \citep[see ][ for the geometry and observational signatures supporting this interpretation]{GP05}.  These components have been identified in other AMSP too \citep{FBP05,FKP05,GP05,FPB07}. The spectral shapes are very similar for individual objects over the course of the outbursts \citep[see e.g.][]{GR98} as well as for different objects \citep{p06}.
  The accretion disc signatures have been also revealed with the \xmm\  data either spectroscopically by the presence of the softer component   as in XTE~J1751--305 \citep{GP05}  or by reduction of the pulse variability amplitude below $\sim$2 keV as in \sax\ \citep{PRA09}. Also the spectral features observed around 6--7 keV in \sax\ are a clear signature of the iron line produced by reflection from the neutral matter,  presumably the accretion disc.

Understanding of the physical nature of the spectral components has important implications for the correct interpretation of the pulse profiles and particularly their strong energy dependence. Phase-resolved spectroscopy reveals that the two major components do not vary in phase \citep{GDB02,GP05}, resulting in prominent soft time lags (i.e. hard photons arriving earlier), which have a steep energy dependence up to 7 keV, the energy where the contribution of the blackbody becomes negligible. The corresponding pulse profiles associated with the components also are significantly different, with the harder photons showing more harmonic content. A natural interpretation of these phenomena is related to the different angular pattern of the blackbody  and the shock emission (PG03).

The variability of the pulse profiles in the course of the outburst also gives us a clue to the origin of the X-ray emission. \sax\ demonstrates remarkably similar evolution of the profiles during its outbursts in  1998, 2002, 2005, and 2008 (H08; \citealt{HPC09}). During most of the outburst (at SD stage, see Fig. \ref{fig:overview}), at high flux level, the profile is very stable and is nearly sinusoidal with a low harmonic content. The profiles observed during the SD stages of the 1998 and 2002 outbursts are almost identical (see  profiles 2 and 4 in fig. 3 in H08).  Also in 2005, profiles 4A and 4B that appeared during a large fraction of the outburst are rather  similar to those in the SD stages of the 1998 and 2002 outbursts.  Even in 2008,  the pulsar  had a similar profile at high flux period (see profiles 1 and 2 in fig. 1 of \citealt{HPC09}). Such profile is consistent with being produced by only one hotspot as discussed and modelled by PG03.  The pulse stability allows to obtain a high photon statistics and use the average pulse profile to get constraints on the neutron star mass-radius relation and the equation of state (see PG03 and  \citealt{LMC08}).

However, there are clear deviations from this profile both at high and low flux levels. In the peak of the outburst in 2002 and 2005,
the profile has a dip in the middle of the broad maximum (see profiles for the P3--P5 period in Fig. \ref{fig:overview}). On the other hand, at low fluxes, the pulse either has a clearly double-peaked profile or a significant skewness opposite to that observed at SD  stage  (H08). This variability also  results in jumps in the phases of the harmonics (timing noise) with the most dramatic example being a glitch-like feature in the phase of the fundamental in 2002 and 2005 outbursts (\citealt{BDM06}; H08).  Such variations seriously complicate the study of the spin evolution of the pulsar. In 1998 this kind of jumps have not been detected as the \xte\ data do not cover well enough the low flux periods (see fig. 1 in  \citealt{GR98} or fig. 1 in H08).

There are plenty of reasons why the pulse profile might change \citep[see][ for a review]{p08}. At high accretion rate, the absorption in the accretion stream might play a role. The effect should be largest close to the phase where the flux is large, when the stream impact point (i.e. the hotspot) is visible at presumably lowest inclination to the normal.  As the accretion rate changes, we expect, for example, some variations of the angular emissivity pattern of the hotspots, and changes in the spot area, their shape, as well as their position at the stellar surface \citep{lamb08, PWvdK09}, as the gas follows different magnetic field lines.   Also, the visibility of the antipodal spot, hidden at high accretion rates,  can vary dramatically as the accretion disc  retreats, causing a strong pulse shape variability.
 
Changes in the accretion disc inner radius are expected on physical grounds as the magnetospheric radius increases with the dropping accretion rate. In addition, there is an indirect evidence for increasing inner radius from the drop of the  kilohertz quasi-periodic oscillation (QPO) frequency during the 2002 outburst  \citep{vSvdKW05}, if interpreted as a signature of Keplerian rotation.

In the present work we track the changes in spectral and timing characteristics through the 2002 outburst of \sax\ with the aim to determine the variations in the geometry of the system.
We first analyse the phase-averaged spectra and their evolution. We study the variations of the apparent hotspot area and the amplitude of the reflection features (which are signatures of cool material in the vicinity of X-ray source) during the outburst. We then study the pulse profiles, their dependence on energy, the corresponding time lags, and finally the phase-resolved spectra.  Furthermore we estimate the physical size of the hotspot and put constraints on the geometry, in particular the inner disc radius and the displacement of the magnetic dipole from the rotational axis. Using  two alternative methods, we determine the magnetic moment of the neutron star.  And finally, we introduce a simple model of the pulsar with two antipodal spots and a variable inner accretion disc radius and compare the predictions of the model to the data.

\section{Observations}

\begin{table}
\caption{Data groupings  for spectral and timing analysis.}
\begin{tabular}{|lll|}
\hline
MJD interval  & Group code  & Outburst stage \\
\hline
52562.13 -- 52563.55 & P &Peak \\
52562.13 -- 52562.46  & P1  &  Peak begins        \\
52562.46 -- 52562.55  &  P2 &         \\
52563.11 -- 52563.14  & P3  &          \\
52563.18 -- 52563.21  &  P4  &          \\
52563.25 -- 52563.55  & P5  &   Peak ends      \\
52564.25 -- 52574.29  & SD & Slow decay\\
52564.17 -- 52564.50 &  SD1 &   Slow decay begins     \\
52564.50 -- 52564.53  &  SD2 &           \\
52565.09 -- 52565.42 &   SD3 &           \\
52565.42 -- 52565.53 &   SD4 &           \\
52566.08 -- 52566.11  &  SD5 &           \\
52566.14 -- 52566.45 &   SD6&         \\
52567.07 -- 52568.42 &  SD7 \\
52569.04 -- 52569.48 &  SD8 \\
52570.03 -- 52570.28 &  SD9 \\
52570.95 -- 52571.39 &  SD10& \\
52571.94 -- 52573.30 &  SD11 & \\
52573.33 -- 52574.29 &  SD12 & Slow decay ends\\
52574.84 -- 52579.36 & RD  & Rapid drop \\
52574.84 -- 52577.40 & RD1 &  \\
52577.40 -- 52579.36 & RD2  & \\
52579.40 -- 52602.79 & FT  & Flaring tail\\
52579.40 -- 52583.30 & FT1 & Flaring tail begins \\
52583.75 -- 52588.44 & FT2 &\\
52588.67 -- 52592.92 & FT3 & \\
52593.02 -- 52602.79 & FT4 & Flaring tail ends\\
\hline
\end{tabular}
\label{t:grouping}
\end{table}

We focus on the data obtained during the most data-rich outburst of \sax\ happened in 2002. The observations are made by the \xte\ during MJD 52562--52604 (October 15 -- November 26) and belong to the ObsID 70080. For the analysis, we used the standard HEADAS 6.1 package and the CALDB. We use the data taken by \xte/PCA (3--20 keV) and HEXTE (25--200 keV); 1 per cent systematics has been added to the PCA spectra (see \citealt{JMR06} for a complete review on PCA calibration). To keep the calibration uniform throughout our dataset, we used the data from PCA units 2 and 3 only.

The outburst can be divided into four periods. The {\em peak} (P) stage (MJD 52562.1--52563.5; 2002 October 15--16) marks the very beginning of the observations, when the 3--20 keV flux (corrected for absorption) is relatively constant at $\sim 2.2 \times 10^{-9}$ erg cm$^{-2}$ s$^{-1}$  and fast (up to $\sim 700$ Hz) QPOs are present. The  {\em slow decay} (SD) stage (MJD 52564.2--52574.3; October 17--27) shows QPOs twice as slow and  the exponential decrease of the flux.  At the {\em rapid drop} (RD) stage (MJD 52574.8--52579.3; October 27 -- November 1),  the flux falls even faster, and the secondary maximum appears in pulse profiles after October 29. During the  {\em flaring tail} (FT), the last stage (MJD 52579.4--52602.8; November 1--24), the flux was rising and fading with the period of  $\sim5$ days till the source become undetectable. The power density spectrum at this stage is dominated by a strong 1 Hz QPO  \citep{vSvdKW05,pwk09}.  The outburst lightcurve and the  sample pulse profiles  are shown on Fig.~\ref{fig:overview} (see also figs 2 and 3 in H08); note the shape differences between various energies.

While the photon count is rather high at the beginning of the outburst, at its end it is necessary to co-add many individual observations to obtain the pulse profiles and spectra with reasonably small errors. Therefore, we group the data  (based on similar count rate and pulse shape) as shown in Table \ref{t:grouping}. For pulse extraction, we used the  pulsar ephemeris obtained by H08.

The spectral analysis (both phase-resolved and phase-averaged) was done using the  \xspec\ 11.2 spectral package \citep{Arn96}.
All uncertainties correspond to a 90 per cent confidence interval.

\section{Spectral analysis}
\label{sec:spectra}

\subsection{Evolution of the spectral shape}

\label{s:ufs}
\begin{figure}
\centerline{\epsfig{file=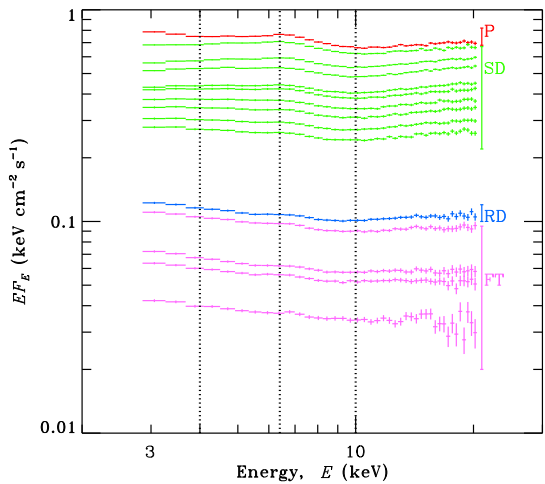,width=7.2cm}}
\caption{The unfolded spectra for \xte/PCA range. Vertical lines indicate the boundaries for spectral components, from left to right: extra flux below 4 keV, 6.4 keV Fe line, blackbody hotspot emission below 10 keV and thermal Comptonization above 10 keV. The outburst stages coloured as in Fig. \ref{fig:overview}  are indicated at the right.}
\label{f:ufs}
\end{figure}

The comparison of the spectral shapes at different observation dates may help to determine the contribution of various emitting components. Thoughout the paper, we plot the so-called unfolded spectra, which show the assumed underlying model multiplied by the ratio of the observed data to the spectral model folded with the detector's response matrix. On Fig.~\ref{f:ufs} we present the \xte/PCA spectra (unfolded using powerlaw model with spectral index 2.0) for several moments of the outburst. In the brightest observations at the peak stage we find the flux excess below 4 keV; this feature is much less noticeable in the later data.
  Most probably the soft excess is the accretion disc emission. This interpretation is supported by the reduced variability at the pulsar frequency below 2 keV in the \xmm\ data obtained during the 2008 outburst \citep{PWvdK09}.  A similar component was found also in the \xmm\ data on XTE~J1751--305 \citep{GP05}. As the disc temperature is below 0.5 keV in these cases, it is not surprising why \xte\ sees the excess only at highest fluxes, when the accretion disc is expected to be hottest.
 
In the 3--10 keV range, the spectrum does not resemble a powerlaw and is likely the blackbody-like emission from the neutron star surface (the similar results were obtained for  other AMSPs, see \citealt*{GDB02,FBP05,FKP05, GP05,FPB07}). Above 10 keV, we observe a powerlaw-like emission with the spectral index slightly changing in time. The hard X-ray data from \xte/HEXTE (see Section \ref{s:spectra}) displays a cutoff at $\sim100$ keV, which is consistent with thermal Comptonization in $\sim50$ keV electron gas. A  bump in the spectrum around 6--7 keV and the hardening above 10 keV are clear signatures of fluorescent iron line at $\sim$6.4 keV and Compton reflection of the underlying continuum from the cold material. The amplitude of the corresponding residuals  reduces as the flux drops.

\subsection{Phase-averaged spectra}
\label{s:spectra}

In this section, we describe the results of fitting the phase-averaged spectra  with various models. All our models include  interstellar absorption,  with the  hydrogen column density fixed at the Galactic absorption value of $1.13\times 10^{21}$ cm$^{-2}$ (obtained from the  HEADAS tool {\sl nh} using the pulsar's coordinates).
The distance to the source is assumed to be $D=3.5$ kpc \citep{gc06}. Because the relative normalization of the PCA and HEXTE instruments is uncertain, we allowed this to be an addition free parameter in all spectral fits.

\begin{figure}
\centerline{\epsfig{file=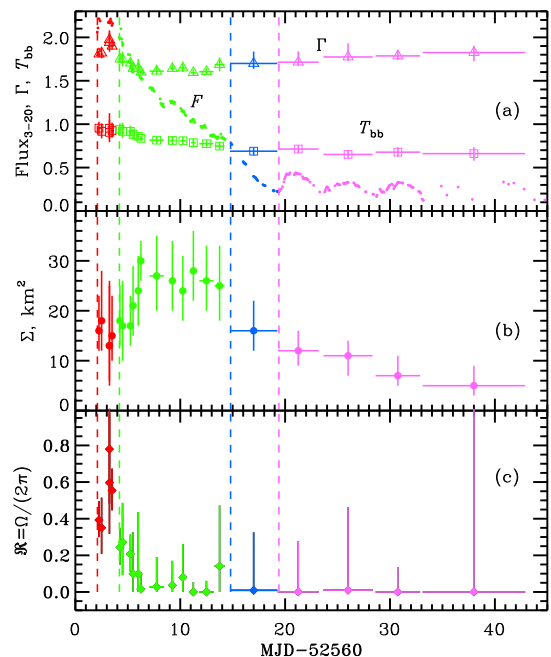,width=7.2cm}}
\caption{The best-fitting parameters for the blackbody and the cutoff powerlaw plus reflection model.
 Evolution of (a) the photon index $\Gamma$ and blackbody temperature $\ktbb$ together with the flux in 3--20 keV band
 corrected for absorption (shown by circles) in units of $10^{-9}$ erg cm$^{-2}$ s$^{-1}$,  (b) blackbody apparent emitting area, and (c)  reflection amplitude. The vertical  dashed lines separate various outburst stages.}
\label{f:pexrav}
\end{figure}

\subsubsection{Powerlaw-based models}
\label{s:pexrav}

A powerlaw model with an exponential cutoff (\xspec\ model {\sc cutoffpl}), described by the photon index $\Gamma$ and cutoff energy $E_\mathrm{cut}$,  produces rather bad fits. Adding a gaussian iron line at 6.4 keV and a blackbody component ({\sc bbodyrad}), described by the temperature $\ktbb$ and normalization $K=[(\rapp/\mathrm{km})/(D/10\ \mathrm{kpc})]^2$, significantly improves the fits giving $\chi^2/{\rm dof}\sim 1$. The apparent residuals remain in the low energy channels below 4 keV,  at the earlier stages of the outburst (groups P1 -- SD4). We model this feature with another blackbody (representing the accretion disc) with the temperature of 0.2 keV and free normalization. The resulting fits show that the temperature of the main blackbody component decreases from about 0.9 to 0.7 keV and the apparent area $\Sigma=\pi \rapp^2$ also decreases from about 30 to 10 km$^2$ as the outburst progresses. At the same time the underlying powerlaw softens from $\Gamma\approx1.5$ to 1.9 and the  cutoff energy increases from about 40 to 80 keV, and during the FT stage it becomes unconstrained.  The amplitude of the additional (disc) blackbody decreases with time, but it is badly constrained as this component is mostly outside of the \xte\ energy band. 

The residuals relative the powerlaw fit seen in Fig.~\ref{f:ufs} imply the presence of the Compton reflection and an iron line in the spectrum. Consequently, we add Compton reflection to the cutoff powerlaw model ({\sc pexrav} model, \citealt{mz95}), with the additional fitting parameters being the reflection amplitude $\refl=\Omega/(2\pi)$ (where $\Omega$ is the solid angle covered by the cold reflector as viewed from the isotropic X-ray source) and inclination which we fix at $i=60\degr$.  The inclination parameter affects only weakly the shape of the Compton reflection continuum in the X-ray band \citep*{mz95,PNS96} and does not affect any other spectral components. Therefore, any other value of $i$ would be also acceptable. 

The limited statistics does not allow to simultaneously constrain $\refl$ and the cutoff energy for the spectra taken only during short time intervals. We have fitted the averaged spectrum for the SD phase and obtained $E_\mathrm{cut}= 63_{-10}^{+13}$ keV; hense we assume $E_\mathrm{cut}=65$ keV in all spectra.  Our results are shown in Fig.~\ref{f:pexrav}. We see that the underlying powerlaw index is almost constant $\Gamma\sim 1.8$. The apparent blackbody area $\Sigma$ reaches the maximum  of $\sim$30 km$^2$ (corresponding to the radius of only 3 km) at the slow decay stage and decreases at the later outburst stages. The reflection amplitude  is clearly larger at the peak and drops rapidly during the SD stage, this explains the apparent softening of the powerlaw seen in the model without reflection.  The absolute amplitude of reflection depends on assumed inclination, scaling roughly as  $\propto 1/\cos i$.

\subsubsection{Thermal Comptonization model}
\label{s:compps}

The physical origin of the powerlaw-like spectrum with a cutoff is most probably related to thermal Comptonization. Because this component is pulsating, it is natural to assume that it is produced in the accretion shock above the neutron star surface. Although the detailed shock structure can be rather complicated, one often approximates it by a single temperature plane-parallel slab. To simulate the Comptonized emission, we use the {\sc compps} model of \cite{PS96} in slab geometry. We set the inclination $i= 60\degr$ following \citet{GDB02}.  As in powerlaw-based models of Sect. \ref{s:pexrav}, inclination is not constrained from the spectral data alone and a particular choice of $i$ will affect only the resulting reflection fraction, which scales as $\propto 1/\cos i$.
The fitting parameters are the temperature and the Thomson optical depth of the Comptonizing electron slab $\kte$ and $\tau$ and  the temperature and the effective emitting area of the blackbody seed photons for Comptonization  $\ktseed$\  and $\shard$. Reflection is described by the  amplitude  $\refl$ and the iron 6.4 keV line ({\sc diskline}) by the normalization.  The statistical limitations of the data   and the energy resolution of PCA  do not allow us to fit the inner disc radius (that controls relativistic smearing of the line and reflection), therefore we fix it (for spectral fits only) on 10$\rg$ (statistically acceptable for all observations, $\rg=2GM/c^2$  is the Schwarzschild radius),\footnote{ Here we assume fixed inner disc radius, because of the limited spectral data. Later in the paper we show that the variations of the reflection amplitude and the pulse profiles require the disc inner radius to change. }
  and assume the radial profile of the emissivity $\propto r^{-3}$. The heated surface around the shock is modeled by an additional blackbody emission component {\sc bbodyrad} \citep[see also][]{GDB02,PG03,GP05, FKP05, FPB07} of temperature $\ktbb$\ and area  $\ssoft$.  To account for the excess flux below 4 keV in groups P1 -- SD4, we used an additional (disc) blackbody component with the temperature of 0.2 keV and free normalization.

\begin{figure}
\centerline{\epsfig{file=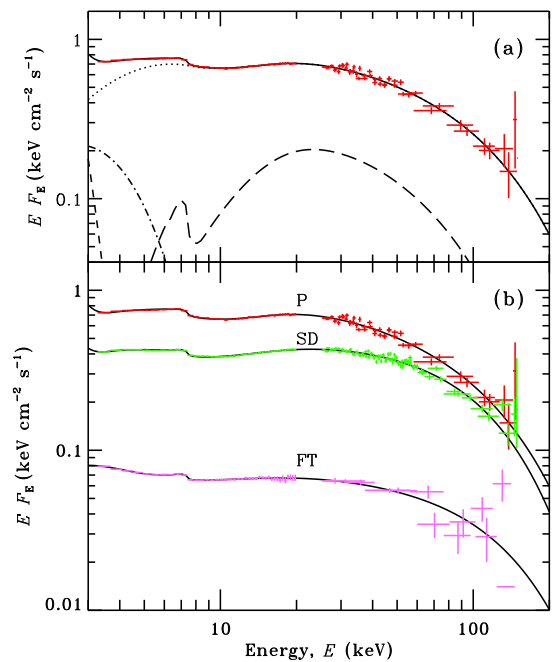,width=7.2cm}}
\caption{Broad-band spectra of \sax\ and the best-fitting thermal Comptonization-based model.
(a) Unfolded observed spectrum at the  peak stage.
Solid, dotted, dot-dashed, long dashed and short-dashed curves represent the total model  spectrum, thermal Comptonization continuum  {\sc compps}, blackbody component, reflection with the 6.4 keV line, and an additional soft blackbody below 4 keV, respectively.
(b) Unfolded observed spectra at the P, SD and FT stages.
Solid curves represent the best-fitting thermal Comptonization model from Sect. \ref{s:compps}.}
\label{f:spectra}
\end{figure}

\begin{figure*}
\centerline{\epsfig{file=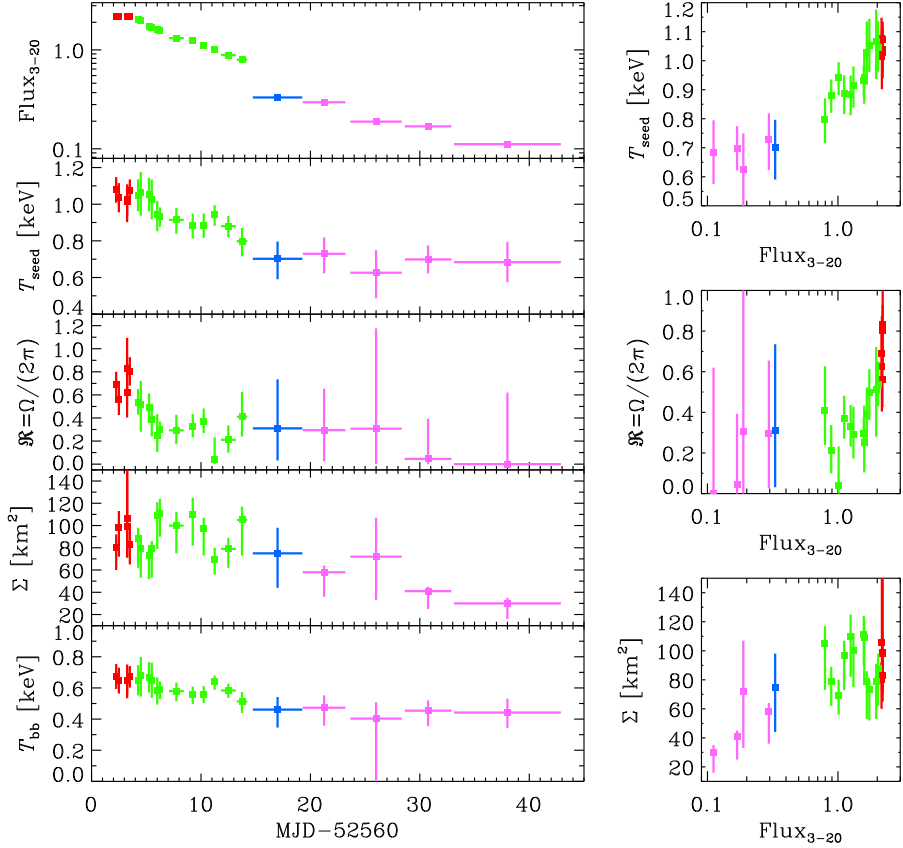,width=12.0cm}}
\caption{The evolution of the spectral parameters for the Comptonization model of Sect. \ref{s:compps} and the correlations between them.
The flux corrected for absorption is in 3--20 keV band  in units of $10^{-9}$ erg cm$^{-2}$ s$^{-1}$.
The points are various outburst stages are coloured as in Fig. \ref{fig:overview}.
}
\label{f:compps}
\end{figure*}

On the physical grounds, we expect that the blackbody emitting area $\ssoft$ is similar to the area covered by the Comptonizing slab $\shard$. Attempts to fit these normalizations independently  lead to unreasonably large effective emitting areas of one of the components (see also \citealt{GP05}, sect. 4.5). We have fitted the spectrum of group SD, which has the best statistic among our dataset, with both $\ssoft$ and $\shard$ being free parameters, and obtained  $\ssoft=75_{-15}^{+31}~\mathrm{km^2}$ and $\shard=30_{-15}^{+30}~ \mathrm{km^2}$. Fixing the ratio  $\ssoft/\shard =2$, gives a good fit and results in a reasonable area of $\shard=37\pm 7~ \mathrm{km^2}$. We then assumed equal areas $\ssoft= \shard= \Sigma$,  and were also able to get good fits  for all data sets and obtained reasonable values for the areas $\Sigma\sim 30$--$110$ km$^2$, which imply the apparent spot radius of $\rapp\sim$3--6 km.  These examples clearly demonstrate that on the quantitative level the results are affected by the assumed relations between the {\sc compps} and {\sc bbodyrad} normalizations.
In the following we assume equal areas  and keep in mind that the systematic error for the spot area is about a factor of two, which translates to a 50 per cent uncertainty in $\rapp$.   We also note that  to obtain the actual spot size at the stellar surface, the apparent areas should be corrected for the inclination and gravitational light bending effect, as discussed in Sect. \ref{s:area}.

The spectral fits to the broad-band spectra averaged over the outburst stages and the  contribution of individual spectral components are shown on Fig.~\ref{f:spectra}. We see that the overall spectral shape does not seem to vary much during the whole outburst as was notices also by \citet{GR98} for the 1998 outburst. However, the fitting parameters do change significantly (see  Fig.~\ref{f:compps}). We clearly see that the seed photon temperature $\ktseed$ as well as the temperature of the additional blackbody are decreasing with time, which is natural as the total luminosity drops. The area $\Sigma$ also seems to decrease.
  This trend is independent of the assumed ratio $\ssoft/\shard$.
Such a behaviour is expected when the magnetospheric radius increases at lower accretion rate. The  values of $\Sigma$ in the Comptonization model are significantly larger than those in the powerlaw model fits (cf. Fig.~\ref{f:pexrav}), because  in
the Comptonization model  only a fraction of the blackbody seed photons escape through the Comptonizing slab and
the Comptonized continuum has a low-energy cutoff unlike  the powerlaw which continues to lower energies without a break. 
Despite the large uncertainties, the reflection $\refl$ shows the decreasing amplitude which is consistent with the trend obtained with {\sc pexrav} model. The Thomson optical depth and the electron temperature of the Comptonizing slab are rather constant, $\tau\sim$1.0--1.2  and  $\kte\approx$40--45 keV, which is consistent with the stability of the underlying powerlaw in the  {\sc pexrav}  model. The decreasing of the reflection together with the drop of the QPO frequencies  (if related to the Keplerian frequencies) is consistent with  the increasing inner disc radius in the course of the outburst.

\section{Temporal properties}
\label{sec:time}

\subsection{Pulse profiles and their energy dependence}
\label{s:sinefits}

The pulse profiles of \sax\ evolve substantially during the 2002 outburst (H08). During the peak stage, the energy-averaged pulse is close to sinusoidal, but skewed to right. The small secondary minimum is noticeable at energies above 10 keV. During the SD stage, the pulse is very stable and nearly symmetric, with a slightly faster rise than decay. In the end of RD1 stage (MJD 52576, October 29), the secondary maximum (signature of the second hotspot) becomes pronounced. At stage RD2  (after MJD 52577, October 30), the pulse becomes skewed to the left, and the secondary maximum shifts to the rising part of the main peak. On MJD 52581 (November 3), the pulse has a clear double-peak profile leaving no doubts that the second hotspot  is visible. At the FT stage, the pulse continues to be skewed to the left. The strength of the first overtone (i.e. at double pulsar frequency) clearly anticorrelates with the flux at later stages as was pointed out  by H08.

\begin{figure*}
\centerline{\epsfig{file=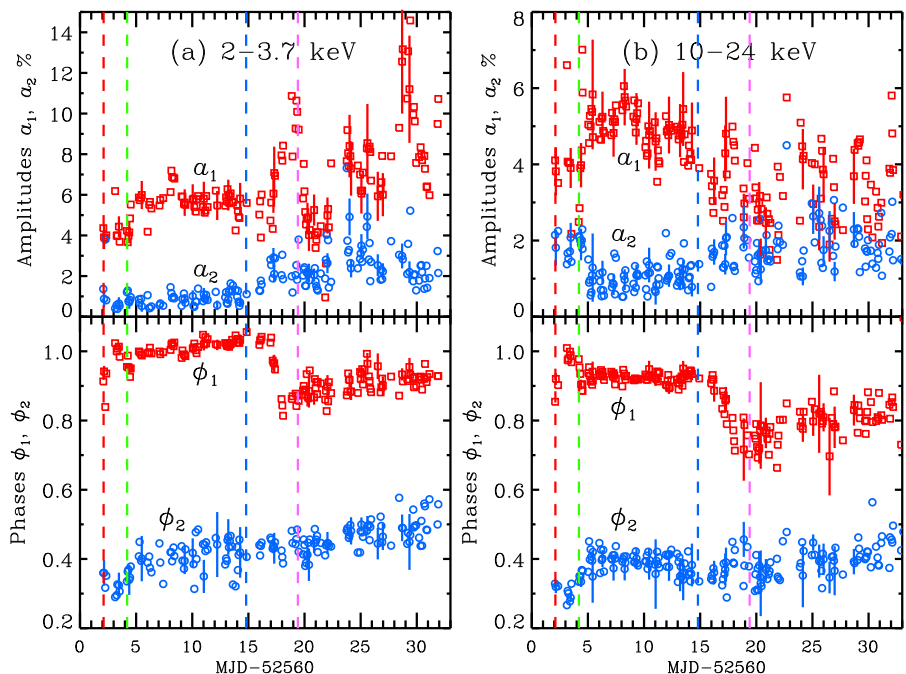,width=12cm}}
\caption{Fourier amplitudes and phases of the best-fitting curves (\ref{eq:cosines}) to the
per-orbit pulse profiles in (a) 2--3.7 keV  and (b) 10--24 keV energy bands.
Top panels: amplitudes of the  fundamental $a_1$ (squares) and the first overtone $a_2$ (circles).
Bottom panels: phases of the fundamental $\phi_1$ (squares) and the overtone $\phi_2$ (circles).
The vertical dashed lines separate various outburst stages. The typical errors are shown for some points.
}
\label{fig:sinefits}
\end{figure*}

We are also interested in the energy dependence of the pulse profiles as this provides us with the clues of the origin of pulse variability.  The evolution during the outburst  of the pulse profiles in two energy intervals  2--3.7 keV and 10--24 keV is presented in Fig.~\ref{fig:overview}.  These energy bands are of particular interest, because the soft one has the largest contribution of the blackbody component, while the hard one contains only the Comptonized continuum. For most of the observations, the pulse shapes can well be described just by two harmonics:
\be
F(\phi)=\overline{F}\{ 1+a_1\cos[2\pi(\phi-\phi_1)] +a_2\cos[4\pi(\phi-\phi_2)] \} .
\label{eq:cosines}
\ee
We have extracted pulse profiles in various energy intervals for every satellite orbit and fitted them with equation (\ref{eq:cosines}). The best-fitting (positive) relative amplitudes $a_1$, $a_2$ and the phases $\phi_1$, $\phi_2$  for the fundamental and first overtone for the two energy intervals are presented in  Fig.~\ref{fig:sinefits}. We also fitted  pulse profiles averaged over the outburst stages  in four different energy band  (see left panels of Fig.~\ref{fig:lag}). We see that the profiles depend strongly on energy.  During the P and SD stages, the relative amplitude of the first overtone is larger at hard energies (above 10 keV). The rms drops with increasing energy only slightly. At the later stages, the rms at energies above 10 keV is 2--3 times smaller than that below 4 keV and the  oscillation amplitude at the fundamental is much more pronounced at soft energies comparing to the hard ones (see also Fig.~\ref{fig:sinefits}).

We also notice that the phase of the overtone behaves differently below 4 keV and above 10 keV.  There is a clear increasing trend at low energies, while the phase is consistent with being constant at higher energies (compare lower panels in Fig.~\ref{fig:sinefits}). This serves as a warning against using such phase-connecting solutions to determine the pulsar frequency evolution \citep[cf.][]{BDM06}. Clearly the pulse profile energy dependence and variability introduces a strong bias.

\subsection{Time lags}
\label{s:lag}

\begin{figure*}
\centerline{\epsfig{file=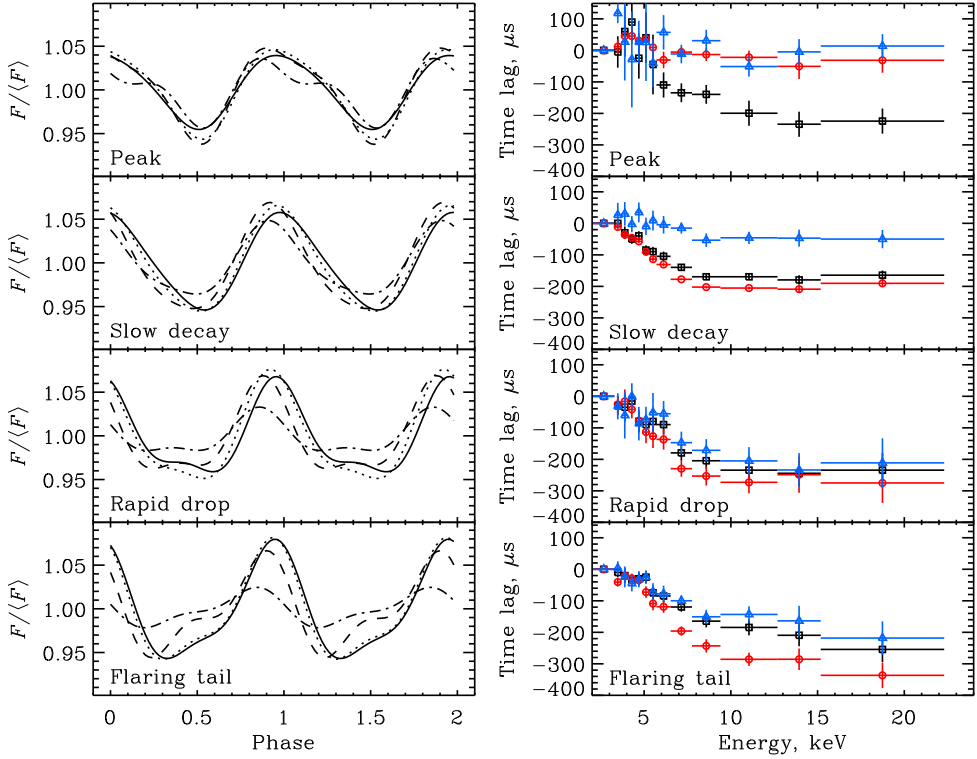,width=13cm}}
\caption{Pulse profiles (left panels) in the  2--3.3, 4.5--4.9, 6.5--7.7 and 15.2--22.3 keV energy bands (solid, dotted, dashed and dash-dotted lines, respectively) as shown by their best-fitting expression  (\ref{eq:cosines}).
 Time lags (right panels) as a function of energy relative to the 2--3.3 energy band for various outburst stages. The time lags obtained from the pulse maximum,  and the lags for the fundamental and the first overtone are shown by black squares, red circles, and  blue triangles, respectively.
}
\label{fig:lag}
\end{figure*}

The pulse profile dependence on energy can also be quantified by studying the time lags. Although they contain less information than the pulse shape, it is customary to study them as the function of energy. In \sax\ the lags are soft, i.e. the pulse peaks at softer energies at a later phase \citep{CMT98,GDB02}.  We compute the time lags by fitting the pulse profiles at a given energy bin with equation  (\ref{eq:cosines}) and
finding the phase difference relative to the reference energy.\footnote{Alternatively, we can use  the standard discrete Fourier transform of the pulse profile to compute the relative amplitudes and phases:
\be \label{eq:fourier}
a_k e^{2\pi i k \phi_k} = 2/N_{\rm ph} \sum_{j=1}^{n} x_j  e^{2\pi i j k/n},
\ee
where $n$ is the number of phase bins, $x_j$ ($j=1,...,n$) is the number of counts in the $j$th bin, $N_{\rm ph}=\sum_{j=1}^{n} x_j $ is the total number of counts, and $k=1, 2$ correspond to the fundamental and the first overtone, respectively. The amplitudes and phases obtained by the two methods become similar in the limit of small errors and only when the profile can be well represented by  equation (\ref{eq:cosines}).} The  90 per cent confidence interval for one parameter is estimated from $\Delta \chi^2=2.71$. For the four  outburst stages, the lags relative to the 2--3.3 keV band  at the fundamental frequency and the first overtone computed from the best-fitting phases $\phi_1$ and $\phi_2$ are  presented in the right panels of Fig.~\ref{fig:lag} by circles and triangles, respectively.  The absolute value of the lags is increasing from 3 up to $\sim$10 keV, and saturates there. Such a behaviour is similar to what is observed in the \sax\ data during the 1998 outburst   \citep{CMT98, GDB02} as well as in other sources \citep[see e.g.][]{GC02,GP05}, except \igr, where the lags seem to change the trend around 15 keV \citep{GM05,FKP05}.  The lags at the fundamental increase when the accretion rate drops (see also \citealt*{HWC08}). They are nearly zero at the peak of the outburst and reach $-300$ $\umu$s at the latter outburst stages. The lags at the first overtone seem to follow similar trends, being consistent with zero at the peak of the outburst and reaching $-200$ $\umu$s latter on.

An alternative measure of the lag is the shift with energy of  the maximum of the best-fitting curve  (\ref{eq:cosines}). To estimate the uncertainties of the lag value, we perturb the fitting parameters within  90 per cent confidence interval and find the maximal displacements of the curve maximum.  The results are shown by squares in  Fig.~\ref{fig:lag}. These lags in general follow the energy dependence obtained from the fundamental, except at the peak stage, where they are significantly larger. This difference originates in the existence  of the absorption-like feature at zero phase in the pulse profile at high energies and the corresponding shift of the maximum to an early phase.

\begin{figure*}
\centerline{\epsfig{file=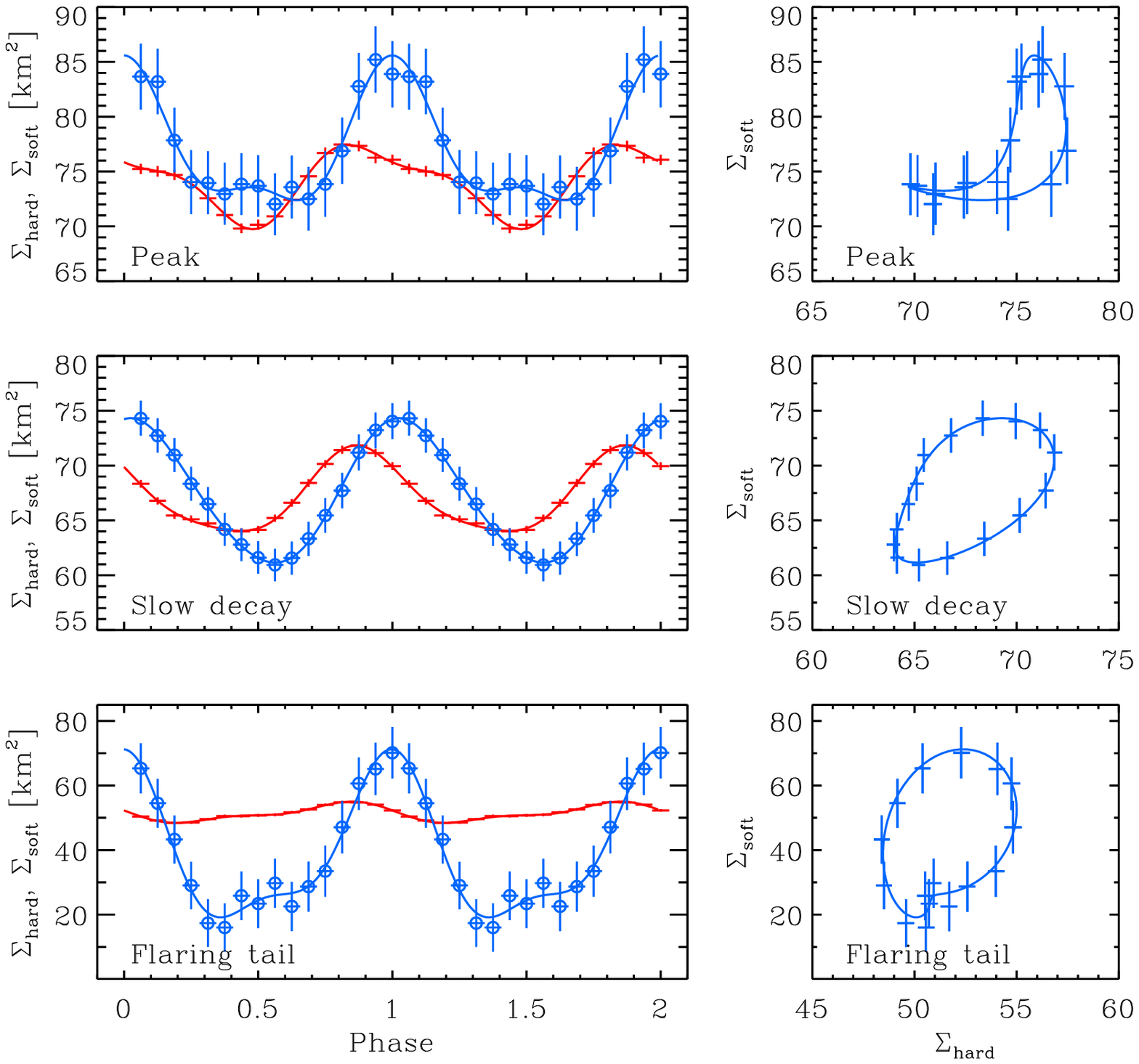,width=11.0cm}}
\caption{Results of the phase-resolved spectral analysis. The best-fitting parameters, except the normalizations of the Comptonization and blackbody components, are frozen at the values obtained for the phase-averaged spectra (groups P, SD and FT).
Left panels: effective emitting area of the two components ($\ssoft$ and $\shard$, blue circles and red crosses, respectively) together with the fits by expression (\ref{eq:cosines}) and parameters given in Table \ref{t:resolve}.
Right panels: $\ssoft$ versus $\shard$.  The outburst stages are indicated on the panels.
}
\label{f:phares}
\end{figure*}

The lags have their origin most probably in difference of the emissivity patterns of blackbody and Comptonization emission, affected, in addition, by the fast stellar rotation (\citealt{GDB02}; PG03). This explains not only why the lag change dramatically below 7 keV, where the  blackbody's   contribution varies significantly, but also the shape of the pulse profile as well as its energy dependence. On the other hand, models involving multiply Compton down- (or up-) scattering \citep[see e.g.][]{CMT98,FT07} pay attention only to the lags ignoring the pulse profiles. A specific time-lag model of \cite{FT07} suffers also from other problems. The soft lags there are explained by multiply scattering of the hotspot's hard radiation in the accretion disc. However, the hard photons above 10 keV are actually mostly reflected (by single Thomson scattering), while the photons at a  few keV are immediately absorbed (by photoelectric absorption) because the disc is rather cool. Thus, the role of multiply scattering is negligible in any case and the lags cannot possibly be produced this way.

\subsection{Phase-resolved spectroscopy}
\label{sec:phaseres}

The phase-resolved spectroscopy of \sax\ for its 1998 outburst has been performed by \citet{GDB02} and for  XTE~1751--305 by \citet{GP05}. It has been concluded, that the energy dependence of the pulse profiles and the soft time lags can be explained with a simple model where  only the normalizations of the Comptonized tail and the blackbody vary. The pulse shapes of these two components are different with the Compton component having stronger harmonic content and peaking at an earlier phase. Such profiles in their turn can be reproduced in a physical model proposed by PG03, where the angular dependence of the emissivity of the spectral components is different. In addition to normalization, one would expect variations of the blackbody temperature caused by Doppler shift, but this is a small effect at the level of 1--2 per cent (PG03).

We have generated the phase-resolved spectra for the P, SD and FT outburst stages. Following \citet{GDB02}, we  used the corresponding thermal Comptonization model fits of the phase-averaged spectra (presented in Section \ref{s:compps}) as a reference. We fixed all the model parameters and fitted the normalizations of the blackbody and thermal Comptonization (denoted as $K_\mathrm{soft}$ and  $K_\mathrm{hard}$, respectively). These normalizations can be directly related to the apparent emitting areas (measured in km$^2$) as  $\Sigma_\mathrm{soft|hard}= \pi D^2 K_\mathrm{soft|hard}$   (where $D=0.35$ is the source distance in units of 10 kpc). The results are shown on Fig.~\ref{f:phares}. The parameters of the fits with  harmonic function (\ref{eq:cosines}) are given in Table \ref{t:resolve}.

\begin{table}
\caption{Harmonic fits by expression (\ref{eq:cosines})
 to the phase-resolved apparent areas.}
\label{t:resolve}
\begin{center}
\begin{tabular}{|lrrrrrr|}
\hline
   & \multicolumn{2}{c}{Peak} &  \multicolumn{2}{c}{Slow decay}  &  \multicolumn{2}{c}{Flaring tail} \\
\hline
Ê  & soft & hard & soft & hard & soft & hard \\
\hline
$\overline{\Sigma}$Ê&  77 & 74ÊÊ& 67Ê& 67Ê& 40Ê& 52 \\
$a_1$			 &  0.08 & 0.04 & 0.10 & 0.06 & 0.60 & 0.05 \\
$\phi_1$  		Ê&  0.01Ê& 0.93 & 0.04 & 0.88 & 0.98 & 0.79    \\
$a_2$  		Ê& 0.03 &  0.02 & 0.01 & 0.01 & 0.21 & 0.02  \\
$\phi_2$  		&  $-$0.01 & 0.27 & $-$0.03 & 0.35 & 0.01 & 0.38 \\
\hline
\end{tabular}
\end{center}
\end{table}

At the peak stage, the blackbody variations are consistent with a sine-wave, while the Comptonized component shows a dip at zero phase, which is reflected in the energy-resolved pulse profiles at high energies  (left panel, Fig. \ref{fig:lag}). This dip (probably associated with absorption in the accretion column) causes the shift of the maximum to earlier phase. The profiles during the SD stage are almost identical to those observed in 1998 \citep{GDB02}, with the hard tail being less variable and having a stronger harmonic. It arrives earlier than the blackbody, resulting in the soft lags. At the FT stage, the blackbody shows enormous variability by a factor of 3, while the rms variability of the Comptonized tail is still only $\sim$5 per cent.  Although  the absolute value for the blackbody emitting area and the amplitude of its variation is model-dependent (because of our assumption of the equal phase-averaged areas and because the $\sim$0.6 keV blackbody contributes little to the \xte\ energy band above $\sim$3 keV), it is clear that the blackbody is much stronger variable than the tail.
These profile shapes are reflected in strong  variations of the energy-resolved pulse profiles shown in Fig. \ref{fig:lag}, with the low-energy pulses (where the blackbody contributes more) having larger rms and the high-energy ones looking exactly as the Comptonized  tail.  The nature of such strong variations of the blackbody is not clear.

\section{Theoretical implications}
\label{sec:discussion}

\subsection{Spot size}
\label{s:area}

The actual spot radius at the neutron star surface can be estimated from the apparent spot area $\Sigma=\pi \rapp^2$ correcting for the light bending and geometry.  Using  \citet{b02} analytical expression for light bending, PG03 obtained the observed bolometric flux produced by a circular blackbody spot of angular radius $\rho$ (visible at all phases) as a function of phase $\phi$ of a slowly rotating pulsar:
\be\label{eq:frho}
F(\phi)=(1-u)^2\!\frac{I_0}{D^2}\pi \rstar^2\sin^2\!\!\rho \Big(Q+ u\tan^2\!\frac{\rho}{2} + U\cos\phi\! \Big) ,
\ee
where $I_0$ is the intrinsic radiation intensity, $\rstar$ is the neutron star radius, $u=\rg/\rstar$ and
\be \label{eq:U}
U  = (1-u)\sin i\ \sin \theta, \quad Q = u+(1-u)\cos i\ \cos \theta,
\ee
and $i$ is the observer inclination and $\theta$ is the magnetic inclination, i.e. the colatitude of the spot centre. The phase-averaged value of $F(\phi)$, corresponding to the expression (\ref{eq:frho}) without the last term in brackets, is related to the apparent radius as
\be
\overline{F} = I \pi \frac{\rapp^2}{D^2},
\ee
where the observed intensity is $I=I_0 (1-u)^2$. Thus for the apparent size of the spot we get
\be
\rapp =R_*\ \sin\rho\ \Big(Q+u \tan^2\!\frac{\rho}{2}\Big)^{1/2},
\ee
which can be reduced to the biquadratic equation for $\sin(\rho/2)$ with the only physical solution:
\be \label{eq:sinrho}
\sin\! \frac{\rho}{2} = \frac{\rapp}{\rstar \sqrt{2} } \left[\! Q\! + \!
\displaystyle
\sqrt{Q^2\!\! -\! \frac{\rapp^2}{\rstar^2}\!\left(1-u\right) \  \cos i\ \cos\theta } \right]^{-1/2} .
\ee
This relation breaks down when parts of the spot are eclipsed, and for the homogeneously bright star we have $\rstar= \rapp/\sqrt{1-u}$ (obviously, then no pulsations can be observed).

\begin{figure}
\begin{center}
\centerline{\epsfig{file=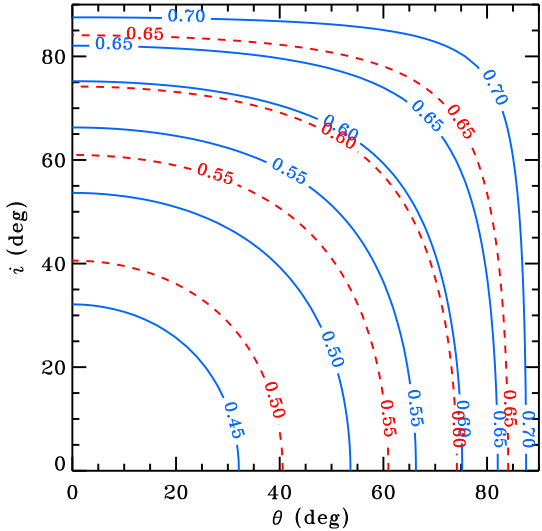,width=7.2cm}}
\end{center}
\caption{Contour plot of constant angular size (in radians) of the emitting spot $\rho$ for the apparent spot radius $\rapp=5$ km. Solid curves correspond to  the neutron star of  $M=1.4\msun$ and radius $\rstar=12$ km. The actual radius of the spot reaches minimum of 5.1 km at $i=0$ and $\theta=0$ and maximum of 8.7 km at $i=90\degr$ and $\theta=90\degr$. The dashed curves correspond to $M=1.7\msun$ and $\rstar=11$ km. In that case, the actual spot size is between 5.1 and 7.6 km.}
\label{f:rho}
\end{figure}

If the apparent size is significantly smaller the stellar radius, there exists a simple relation between the physical and the apparent sizes (PG03):
\be \label{eq:rhostar_Q}
\rho\rstar= \rapp \  Q^{-1/2}.
\ee
The smallest possible spot radius is close to the apparent radius as $Q=1$ for $i=0$ and $\theta=0$. As the oscillation amplitude is small, we expect that $\theta$ is rather small, thus the reasonable upper limit on the spot size is reached at $i=90\degr$ (taking $\theta=0$):
\be
\rho\rstar=  \rapp / \sqrt{u}.
\ee

Using equation (\ref{eq:sinrho}) we compute the dependence of $\rho$ on $i$ and $\theta$ for a given $\rapp$. The results for  $\rapp=5$ km (corresponding to the apparent area of 80 km$^2$) and two stellar compactnesses are shown in Fig. \ref{f:rho}.  The actual spot radius  $\rho\rstar$ is about 5--8 km depending on the inclination and compactness.  The effect of rapid rotation changes the results only slightly.  However, the apparent spot size obtained from the fitting of the spectra of \sax\ is model dependent. The powerlaw-based model (Fig. \ref{f:pexrav}) gives an apparent radius $\rapp$ between 3 and 1 km (for the assumed distance of $D=3.5$ kpc).  The radius estimated from the thermal Comptonization models  (Fig. \ref{f:compps})  is significantly larger, varying from about 5.6 to 3.6 km (corresponding to $\Sigma=$100--40 km$^2$) during the outburst. The main reason for this discrepancy is that the powerlaw continues to lower energies without a cutoff, while thermal Comptonization turns over at energies corresponding to seed photons. In addition, for Thomson optical depth of the order unity in the Comptonizing slab, only about 1/3 of the seed photons can escape directly, the rest being scattered and Comptonized to higher energies. Thus on the physical grounds we prefer the larger apparent size.

The apparent area is still uncertain within a factor of two as we discuss in Sect. \ref{s:compps}, because of the unknown ratio of the areas of the blackbody and Comptonized components.  The corresponding 50 per cent uncertainty in $\rapp$ means that the actual spot sizes of course are also uncertain by about the same amount, as it is clear from equation (\ref{eq:rhostar_Q}). An additional complication comes from the deviation of the spectrum from the blackbody. The actual spot size can be larger by the colour correction factor $\fcol=T_{\rm col}/T_{\rm eff}$, which describes the shift of the spectral peak relative to the blackbody. However, for the atmospheres heated from above, this correction should not play a significant role (see discussion in PG03). In that case,  the spot size during the whole outburst is smaller than the stellar radius  for the typical neutron star masses and radii.

\subsection{Oscillation amplitude and constraints on the geometry}
\label{s:geom}

\begin{figure*}
\begin{center}
\leavevmode \epsfxsize=7.0cm \epsfbox{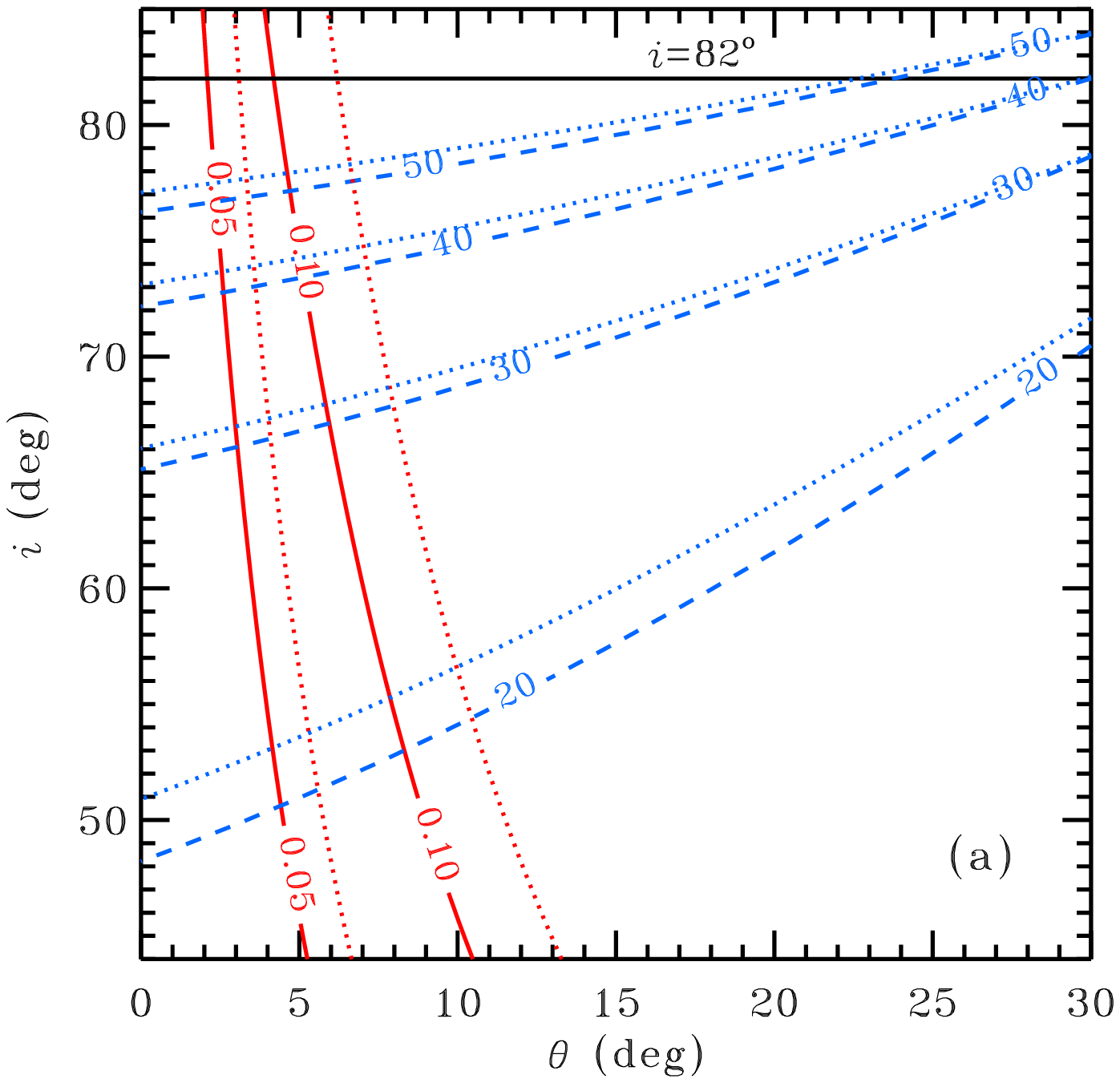} \hspace{1cm}
\epsfxsize=7.0cm \epsfbox{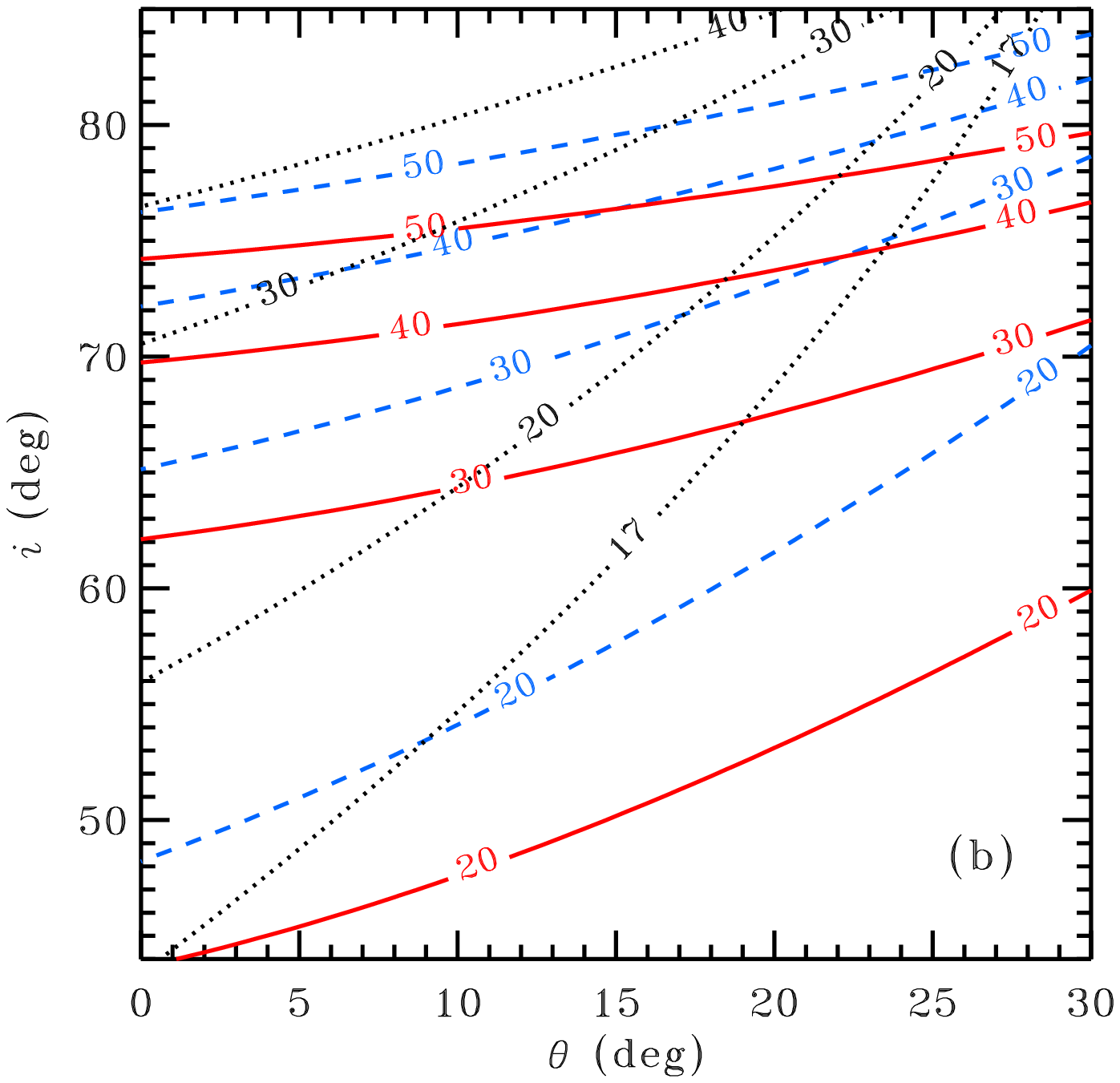}
\end{center}
\caption{
 Constraints on inclination and colatitude of the emitting spot.
(a) The solid (nearly vertical) curves show the contours of constant oscillation amplitude $a_1$ (given by equation [\ref{eq:ampl_spot}]) of 5 and 10 per cent expected from a blackbody spot at a neutron star of mass  $M=1.4\msun$ and radius $\rstar=12$ km. The spot size corresponding to $\rapp=5$ km is computed using equation (\ref{eq:sinrho}), see Fig. \ref{f:rho}.
The dashed curves correspond to the upper limit on the inner disc radius (in km) that is required to fully block the view of the antipodal emitting spot.  The corresponding results for the neutron star of $M=1.7\msun$ and $\rstar=11$ km are shown by dotted curves.
The upper limit on the inclination of 82\degr  \citep{bc01} is shown by the horizontal line.
(b) The inner disc radii required to block the antipodal spot ($M=1.4\msun$ and $\rstar=12$ km) for various apparent spot sizes $\rapp=3$ (solid curves), 5 (dashed), and 7 km (dotted). 
}
\label{f:screen}
\end{figure*}

Some constraints on the system geometry can be obtained from the observed oscillation amplitude. For a blackbody emitting spot at a slowly rotating star, the relative amplitude $a_1$ of the fundamental is (see PG03 and equation [\ref{eq:frho}]):
\be\label{eq:ampl_spot}
a_1 =\frac{U}{Q+u\tan^2(\rho/2)}.
\ee
Comparing this expression to the observed oscillation amplitudes, we can constrain the geometrical parameters $i$ and $\theta$.
The spot angular size $\rho$ is estimated from equation (\ref{eq:sinrho}) and substituted to  equation (\ref{eq:ampl_spot}). The resulting constraints on  $i$ and $\theta$ are shown in Fig.  \ref{f:screen}(a) as the contours of constant amplitude $a_1$ of 5 and 10 per cent. We see that for most probable inclinations, the displacement of the spot centre from the rotational axis is just 3\degr--7\degr. These results only weakly depend on the assumed stellar compactness and stellar rotation \citep{PB06}. A stronger effect comes from the deviation of the emissivity pattern from the blackbody, as the Comptonization tail is expected to have a broader beaming pattern and therefore smaller amplitude (PG03, \citealt{VP04}). The same observed $a_1$ requires a slightly larger $\theta$ for a given $i$ \citep[see][]{GP05}, but still $\theta\lesssim 10\degr$. Inclination smaller than about 40\degr  (and thus possibly larger $\theta$) can be ruled out on the basic of the X-ray \citep{CM98} and optical \citep*{GHG99, HCC01,WCR01,DHT08} orbital modulations.

\subsection{Visibility of the antipodal spot and the inner disc radius}
\label{s:innerdisc}

The simplicity of the sine-like pulse profiles observed during the SD  stage can be used to argue that the antipodal spot is not visible at this stage. The appearance of the antipodal spot later in the outburst is natural as the accretion disc is expected to recede from the neutron star opening the view of the lower stellar hemisphere. If we associate the appearance of the secondary peak in the pulse profile  (on 2002 October 29) with the appearance of the antipodal spot, we can estimate the disc radius at this specific moment. This then can be used to get independent constraints on the magnetic field. The results depend somewhat on the spot size. We first assume $\rapp=5$ km (corresponding to apparent area of 80 km$^2$, see Fig. \ref{f:compps}), and then vary it by 50 per cent within the uncertainty range.

Let us consider  the parameter space where the antipodal spot can be eclipsed by the disc. As parameters we take the inclination $i$ and the spot centroid colatitude $\theta$. We take the spot size $\rho$ given by equation (\ref{eq:sinrho}) and shown in Fig. \ref{f:rho} and numerically compute photon trajectories from the spot elements  towards the observer at every pulsar phase and check whether it crosses the disc (see Appendix \ref{app:discabs}). This allows us to estimate the maximally allowed inner disc radius $\rin$, which fully blocks the spot. The contours of $\rin$ at the plane $i$--$\theta$  for $\rapp=5$ km are shown in Fig. \ref{f:screen}(a). We see that at large inclinations $i\sim80\degr$, the disc fully covers the antipodal spot even when the inner radius is rather large $\rin\sim40$ km. At $i\sim50\degr$, the disc has to extend to $<20$ km to cover the spot. The disc extending to the corotation radius
\be \label{eq:rco}
\rco= 31 \left( \frac{M}{1.4\msun} \right) ^{1/3}  \left( \frac{\nu}{401\ \mathrm{Hz}} \right) ^{-2/3} \ \mathrm{km}
\ee
blocks the antipodal spot from any observer at $i\gtrsim67\degr$ (for $M=1.4\msun$ and $\rstar=12$ km).
 
If we assume a smaller spot $\rapp=3$ km, the inner disc radius needed to fully block the spot increases by about 20 per cent
compared to the $\rapp=5$ km case (compare solid to dashed curves in Fig. \ref{f:screen}b). On the other hand, increasing the apparent spot size to  7 km, necessarily decreases the disc radius by  20 per cent (see dotted contours). The stellar compactness influences the results very little, for example, taking a higher mass of $M=1.7\msun$ and $\rstar=11$ km, lead to the decrease in the disc radius by only $\sim$1 km (compare dashed and solid curves in Fig. \ref{f:screen}a).

If the accretion is centrifugally inhibited at $\rin>\rco$, the clear signatures of the antipodal spot during the FT stage, give us an upper limit on the inclination $i\lesssim64\degr, 67\degr, 73\degr$ for $\rapp=3, 5, 7$ km, respectively (for $M=1.4\msun$), which is lower than the constraint $i<82\degr$ obtained from the absence of binary eclipses \citep{bc01}.  For a more massive star $M=1.7\msun$, the corresponding limits on the inclinations increase by 3 degrees. If, however, the inclination actually is $\gtrsim75\degr$,  this would imply that accretion proceeds even when $\rin>\rco$ \citep*[see ][]{ST93,RFS04,KR07}.

\begin{table}
\caption{Estimation of the inner disc radius and stellar magnetic moment $\mu$
 from the apparent spot area at 2002 October 29.
}
\centering
\begin{tabular}{lcccccc}
\hline
 & \multicolumn{3}{c}{$M=1.4\msun$}   &  \multicolumn{3}{c}{$M=1.7\msun$}  \\
 & \multicolumn{3}{c}{$\rstar=12$ km}   &   \multicolumn{3}{c}{$\rstar=11$ km}  \\
\hline
$i$ (deg) & 50 & 60 & 70  &  50 & 60 & 70    \\
\hline
&\multicolumn{6}{c}{$\rapp=3$ km} \\
$\rin$ (km)   & 21.5 & 27.5 & 40   & 20.5 & 27 & 38  \\
$\mu\ \ka^{7/4}f_{\rm ang}^{-1/2}$ ($10^{25}\mathrm{G\ cm}^{3}$)   & 4.4 & 6.8 & 13.2  & 3.6 & 5.9 & 10.7  \\
\hline
&\multicolumn{6}{c}{$\rapp=5$ km}  \\
$\rin$ (km)   & 19.5 & 24 & 34   & 18 & 23 & 33  \\
$\mu\ \ka^{7/4}f_{\rm ang}^{-1/2}$ ($10^{25}\mathrm{G\ cm}^{3}$)   & 3.7 & 5.4 & 9.9  & 2.9 & 4.5 & 8.4  \\
\hline
&\multicolumn{6}{c}{$\rapp=7$ km} \\
$\rin$ (km)   & 17 & 20 & 27  & 14 & 18 & 26  \\
$\mu\ \ka^{7/4}f_{\rm ang}^{-1/2}$ ($10^{25}\mathrm{G\ cm}^{3}$)   & 2.9 & 3.9 & 6.6  & 1.9 & 2.9 & 5.5  \\
\hline
\end{tabular}
\label{tab:rin}
\end{table}

The  constraints on the oscillation amplitudes (see Sect. \ref{s:geom} and solid curves in Fig. \ref{f:screen}a)  reduce dramatically the allowed region in the $(i,\theta)$ plane. Combining this together with the fact that the antipodal spot has just appeared to the view, gives directly the estimation on the inner disc radius at that moment. The results are presented in Table \ref{tab:rin} for various inclinations  and two sets of stellar compactness. We see that the largest uncertainty comes from the unknown inclination. The compactness affects the results at less than 10 per cent level, while  a factor of 2 error in the apparent spot area (i.e. 50 per cent error in $\rapp$) results in only a 20 per cent error in $\rin$.

\subsection{Constrains on stellar magnetic field}
\label{s:magn}

Having the information about the inner disc radius allows us to estimate the stellar magnetic moment. We need to know the accretion rate, which can be obtained from the observed bolometric flux $F$:
\be \label{eq:LFfang}
L =4\pi D^2 F f_{\rm ang}=\eta \dot{M} c^2 ,
\ee
where $L$ is the total luminosity, $\eta=1-\sqrt{1-u}$ is the accretion efficiency in Schwarzschild metric, and $f_{\rm ang}$ is the factor correcting for the angular anisotropy of radiation and general relativity effects. It  depends on the inclination, stellar compactness and spin, as well as position of the emitting spot and its emission pattern (see Appendix \ref{app:flux} for details of calculations).
We approximate the angular distribution of radiation from the spot by a law
\be \label{eq:intani}
I_*(\alpha)=I_0 (1+\ani\cos\alpha),
\ee
where $\alpha$ is the angle relative to the stellar normal and $\ani$ is the anisotropy parameter. The value of $\ani=0$ corresponds to the blackbody-like emission pattern and the negative $\ani$ correspond to the Comptonized emission from an optically thin slab (see PG03; \citealt{VP04}).  For inclinations in the range 50\degr--70\degr and a circular spot, $f_{\rm ang}$ varies in a rather narrow range $1.3\pm0.2$ (see Fig. \ref{f:fang}). It depends very little on the assumed spot size, when $\rapp$ varies from 3 to 7 km, this factor changes only by 2--3 per cent.

The inner disc radius depends on the accretion rate $\dot{M}$ and scales with the classical Alfv\'en radius as
\be \label{eq:alfven}
\rin =\ka (2GM)^{-1/7} \dot{M}^{-2/7} \mu^{4/7} ,
\ee
where $\mu$ is the neutron star magnetic moment. The coefficient $\ka$ is rather uncertain, with the numerical simulations giving $\ka=1/2$ \citep*{long05}.  Thus, we finally can express the magnetic dipole moment  as
\beq\label{eq:magmom}
\mu_{25} & =&  0.56\times  \ka^{-7/4}
\ \left(\! \frac{M}{1.4\msun} \right)^{1/4}
\left(\!\frac{\rin}{10\ \mathrm{km}} \right)^{7/4}
\nonumber \\
&\times&  \left(\!  \frac{f_{\rm ang}}{\eta} \frac{F}{10^{-9}\ \mathrm{erg}\ \mathrm{cm}^{-2}\ \mathrm{s}^{-1}} \!  \right)^{1/2} \!\! \frac{D}{3.5\ \mathrm{kpc}}\  ,
\eeq
where $\mu_{25}=\mu/10^{25}\ \mathrm{G}\ \mathrm{cm}^3$.

The antipodal spot appears after 2002 October 29 at the flux level in the 3--20 keV band of $F_{3-20}\approx 0.4\times10^{-9}$ erg cm$^{-2}$ s$^{-1}$ and the corresponding bolometric flux $F \approx 0.8\times 10^{-9}$ erg cm$^{-2}$ s$^{-1}$ and luminosity of about $L=1.1\times 10^{36}$ erg s$^{-1}$. Substituting this to equation (\ref{eq:magmom}), we get the magnetic dipole moments, which  are presented in Table \ref{tab:rin} for two sets of stellar parameters and various inclinations.  Rather conservative limits  are $\mu_{25}\approx (9\pm5)\ \ka^{-7/4}$ (for anisotropy factor $f_{\rm ang}=1.3\pm0.2$), with the largest uncertainty coming from the unknown inclination and  parameter $\ka$. 
The estimated surface magnetic dipole field is about $B_0\approx (0.8\pm0.5)\times 10^8\ \ka^{-7/4}$ G  (assuming neutron star radii 10--15 km), which is in excellent agreement with the value obtained recently from the pulsar  long-term spin-down rate in the quiescence (H08), if $\ka\sim 1$.  A similar value (close to the upper end) can be obtained \citep[see][]{GR98} assuming that a sharp break (at 2002 October 28) in the pulsar light curve is associated with the onset of the propeller effect \citep{IS75}.

On the other hand, an independent knowledge of the stellar magnetic field  provides constraints on parameter $\ka$. Taking $\mu_{25}=5\pm3$ as obtained by H08, together with our estimations of the inner disc radius from Table \ref{tab:rin}, we get very  conservative limits (for inclinations $50\degr<i<70\degr$)
\be
0.4 \lesssim \ka \lesssim 2.5 .
\ee
   These limits include all systematic and statistical  uncertainties, including uncertainties in inclination, stellar mass, anisotropy factor, spot size and magnetic moment.
As all theoretical models predict $\ka<2^{1/7}\approx1.1$ \citep[see discussion in][]{PC99,KR07}, this limits the
inner disc radius to $\rin \lesssim31$ km and gives some constraints on the inclination $i \lesssim 75\degr$.

\begin{figure}
\begin{center}
\centerline{\epsfig{file=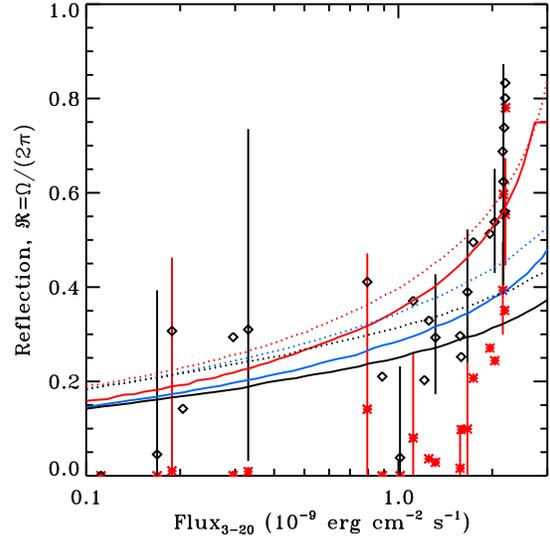,width=7.2cm}}
\end{center}
\caption{Reflection amplitude $\refl=\Omega/(2\pi)$ as a function of the flux in 3--20 keV band.
  The data points shown by diamonds  and stars (with representative error bars) are for the fits using thermal Comptonization model of Section \ref{s:compps}  and powerlaw-based model of Section \ref{s:pexrav}, respectively.
The set of three curves give the theoretical dependence expected for three different geometries of the emitting spot:
(from top to the bottom)  a narrow ring with the outer angular radius $\rho_{\max}$
defined by the dipole formula (\ref{eq:dipole}) and the inner angular radius $\rho_{\min}=0.9\rho_{\max}$;
a  ring with inner angular radius $\rho_{\min}=\rho_{\max}/2$;
and a circle of radius $\rho_{\max}$ centred at the rotational pole. 
The solid curves are for the black body emitting spot (anisotropy parameter $\ani=0$), and
the dashed curves correspond to the angular pattern of the Comptonized emission with $\ani =-0.7$,
with the magnetic moments of $\mu_{25} =6$ and 7, respectively, and $\ka=1$.
Neutron star mass is $M=1.4\msun$ and inclination $i=60\degr$.
}
\label{f:reflect}
\end{figure}

\begin{figure*}
\centerline{\epsfig{file=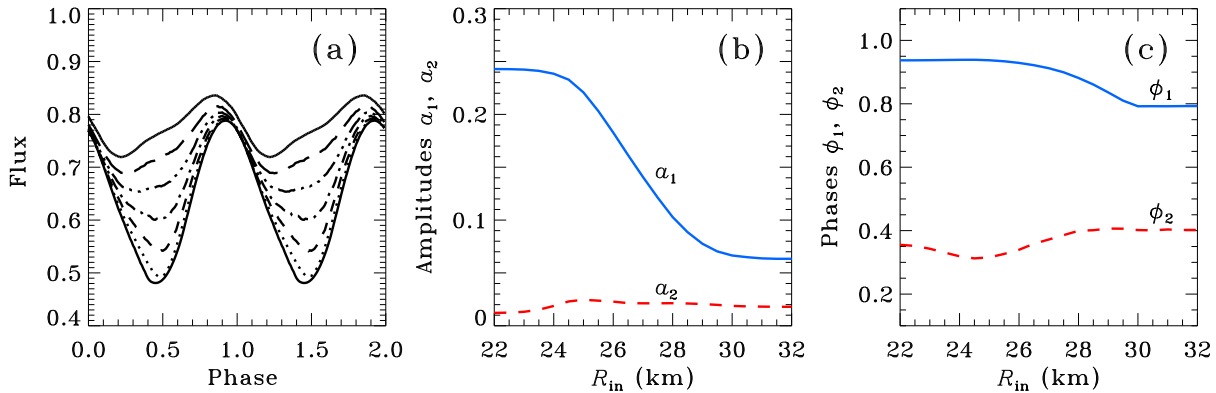,width=16cm}}
\caption{(a) Pulse profiles from two antipodal spots with the inner disc radius $\rin$ varying from 24 to 30 km (from bottom to top).  At $\rin=24$ km, the view of the antipodal spot is fully blocked by the disc, while for $\rin=30$ km the whole spot is visible.
(b) The amplitudes and (c) the  Fourier phases of the fundamental and first overtone of the pulse profiles, computed using equation (\ref{eq:fourier}),  as a function of $\rin$. Parameters of the simulations: neutron star mass $M=1.4\msun$ and radius $\rstar=2.5\rg=10.3$ km, inclination $i=65\degr$, circular blackbody spots  of angular radius $\rho=$30\degr centred at a colatitude $\theta=16\degr$.
}
\label{fig:spot_disc}
\end{figure*}

\subsection{Varying reflection}
\label{s:refl}

Variation of the reflection amplitude  $\refl$ during the outburst and its correlation with the flux (see Figs \ref{f:pexrav} and \ref{f:compps}) are clear signatures of the changes in the geometry happening when the accretion rate decreases. At the peak stage, our thermal Comptonization fits require reflection to be $\refl\sim$ 0.6--0.8
  and powerlaw-based models give $\refl\sim$ 0.4--0.6. 
 The reflection substantially decreases during the SD phase and is not constrained at the FT stage. Although the actual reflection amplitude is model dependent (depends on the model of the underlying continuum and inclination), the trend is not. The dramatic changes in $\refl$ can only be explained if the geometry of the accretion flow varies significantly and  the inner radius of the accretion disc in the peak of the outburst is close to the neutron star surface.
Assuming $\rin$ in the range 12--15 km at the peak flux level of $F_{3-20}\approx 2\times10^{-9}$ erg cm$^{-2}$ s$^{-1}$, we can get a rough estimate of the stellar magnetic moment using equation (\ref{eq:magmom}). Taking anisotropy factor $f_{\rm ang}=1.3\pm0.2$ (see Fig. \ref{f:fang}) and stellar masses in 1.4--1.7$\msun$ range, we get $\mu_{25}=(6\pm 2)\ka^{-7/4}$, where the largest uncertainty comes from a strong dependence on $\rin$. Such $\mu$ is consistent with the estimation from the antipodal spot eclipse in Table~\ref{tab:rin}.

  For illustration, we simulate the expected dependence of the reflection amplitude on flux.  For  various $\ani$, we compute the flux $F$ as a function of $\dot{M}$ at a given inclination $i$ (as described in Appendix \ref{app:flux}). The reflection amplitude $\refl$ of course depends on the inner disc radius $\rin$ (see details of calculations in Appendix \ref{app:refl}), which varies with $\dot{M}$ according to equation (\ref{eq:alfven}).
Both flux and reflection depend on the geometry of the emitting spot.  As the constraints obtained in Section \ref{s:geom} imply a rather small displacement of the spot centroid from the rotational axis,  for estimating the  reflection, we can neglect it and assume that magnetic dipole is aligned with the rotational axis.
  We consider three geometrical arrangements: a full circle centred at the rotational pole, a ring-shaped spot with the inner angular radius equal half of the outer radius and a very narrow ring. In all cases, we take the maximum spot extend from the rotation axis $\rho_{\max}$ given by the dipole formula:
\be \label{eq:dipole}
\sin^2 \rho_{\max}= \frac{\rstar}{\rin} .
\ee
  
The theoretical reflection-flux dependence are shown in Fig. \ref{f:reflect} together with the data.  We see that the circular spot does not reproduce a sharp dependence of $\refl$ on flux. On the other hand, a narrow ring is more consistent with the data, as in that case at high fluxes, when the inner disc is close the star, the emission region is closer to the equator giving a strong increase in the reflection fraction.  A sharp increase of reflection and its high value at high fluxes, can also be explained if the inner disc is puffed up \citep[see e.g.][]{long05} because of the interaction with the magnetosphere.

\subsection{Pulse profile variations with receding disc}
\label{s:disc}

 As we argued above, the decreasing area of the emitting spot $\Sigma$ and of the reflection amplitude  $\refl$ when the flux is dropping are strong evidences of the receding disc. This is natural as the magnetosphere radius is expected to increase.
Let us now look at the pulse profile variation in the course of the outburst and compare to the expected behaviour.

When the accretion rate is high, the disc should be close to the star and may fully blocks the view of the antipodal spot as we discussed in Section \ref{s:innerdisc}. The accretion flow itself can be optically thick and emission from the hotspot can be partially absorbed by the flow. Using the mass conservation equation $\dot{M}= 4 \pi \rstar^2 f V \rho$ (where $f\sim 0.1$ is a  fraction of the stellar surface covered by the accretion stream, which corresponds to the spot area of about 100 km$^2$)  and taking accretion velocity of $V\approx c/3$, we get the Thomson optical depth  through the accretion stream at a typical distance $\rstar$ from the star:
\be\label{eq:tauflow}
\taut \approx \rho \rstar \frac{\sigmat}{\mpro} = 0.3 \frac{\dot{M}}{10^{16} \mathrm{g\ s}^{-1}} \frac{0.1}{f},
\ee
where $\sigmat$ is the Thomson cross-section, $\mpro$ is the proton mass. At the peak of the outburst, the bolometric flux is $\sim4 \times 10^{-9}$ erg cm$^{-2}$ s$^{-1}$, and luminosity $\sim 6\times 10^{36}$ erg s$^{-1}$, which gives the accretion rate of about $2 \times 10^{16}$ g s$^{-1}$ and $\taut\sim 0.6$. Thus, we see that the peak of the outburst the flow is marginally optically thick and can produce dips in the light curves blocking the hotspot, while at lower fluxes corresponding to most of the outburst, the flow is transparent.  It is possible that this effect is  the cause of the observed non-sinusoidal pulse shape at the peak of the outburst  (see the inset for P3--P5 in Fig.~\ref{fig:overview} and upper left panel in Fig.~\ref{fig:lag}).  The effect is clearly largest at hard energies, which might imply that the hard Comptonized radiation is coming from a smaller area (which is easier to block) that the softer blackbody radiation.

As we discussed in the Introduction, the pulse profile during SD stage is very stable and is similar to the profiles observed during other outbursts at similar flux level. The profiles are well described by a one-spot model, implying that the disc is still rather close to the star blocking the view of the antipodal spot. The dependence of the profiles on energy is consistent with the being produced by the different emissivity pattern of the blackbody and Comptonized radiation modified by Doppler effect (PG03). The pulse stability implies that the position of the spot at the neutron star surface does not change significantly. During the SD stage, the flux drops by a factor of 3, which corresponds to an increase of the inner disc radius by 40 per cent and corresponding increase of the outer spot angular radius by 20 per cent according to equations (\ref{eq:alfven}) and (\ref{eq:dipole}). However, these variations seem to have little impact on the pulse profile, which is expected if the centroid of the spot does not move and the spot size is smaller than the stellar radius \citep{PB06}.

With the dropping accretion rate, the disc moves sufficiently far so that the observer can see at least part of the antipodal pole. We thus expect that the pulse profiles change correspondently.  Significant variations in the pulse profile are observed from the middle of the RD phase.  These changes are also reflected in sharp jumps in the phase of the fundamental (see Fig. \ref{fig:sinefits}).  This motivates us to model theoretically variations of the pulse profiles with the varying inner disc radius.

As an illustration, we consider two circular antipodal spots emitting as a blackbody and displaced from the rotational axis by angle $\theta$. The light curve produced by each spot is computed following techniques described in PG03 and \citet{PB06}, accounting for Doppler boosting, time delays, and gravitational light bending. Now we also account for the effect of absorption by the disc of varying inner radius $\rin$. We compute the trajectory for photons emitted by each spot element at every phase and check whether it crosses the disc plane at radius smaller or larger than $\rin$ (see Appendix \ref{app:discabs} for details). Fig. \ref{fig:spot_disc}(a) shows the pulse profile variations as a function of the inner disc radius. Noticeable signatures of the  second spot appear when $\rin>27$ km (for $i=65\degr$),  the dramatic change in the pulse profile happens when the inner radius changes  from 28 to 30 km, as most of the antipodal spot then appears to our view. These fast changes are also reflected in a rather large  phase shift $\Delta\phi_1\approx0.2$  of the fundamental, but small variations of the phase of the overtone, as shown in  Fig. \ref{fig:spot_disc}(c).    This behaviour is qualitatively similar to that  observed  in \sax\ (see Figs \ref{fig:sinefits} and \ref{fig:lag}). 
Thus the phase jumps can be explained by changing visibility of the antipodal spot as the accretion disc recedes from the neutron star. The timing noise observed in other pulsars \citep{pap07,pap08,RDB08} may also have its origin in this effect. In addition,    spot wandering \citep{lamb08, PWvdK09} and variations in the spot size, shape, and emissivity pattern  might be involved. 

We also show the variation of the amplitudes in  Fig. \ref{fig:spot_disc}(b). We see that the fundamental amplitude decreases smoothly as $\rin$ grows\footnote{We note that for a slowly rotating star, two antipodal blackbody spots which are visible all the time produce no pulsations whatsoever \citep[see e.g.][]{b02,PB06}. Thus it is not surprising that even for a star rotating at 400 Hz, the amplitude of the fundamental decreases when both spots become visible.}, while the amplitude of the first overtone stays nearly constant. Such a behaviour is consistent with that observed in Fig. \ref{fig:sinefits}(b) (see also fig. 1 in \citealt{HPC09}), but is not consistent with that seen in Fig. \ref{fig:sinefits}(a). We note, however, that in real situation the spots are probably not circular and emission is not a blackbody.

\section{Summary}

\label{sec:conclusions}

In this paper, we present the detailed analysis of the 2002 outburst data of SAX1808. Below  we summarize our findings.

\begin{enumerate}

\item
The phase-averaged spectra of the pulsar are roughly described by a powerlaw with the photon index $\sim$2. However,  additional features are significantly detected: iron line at around 6.4 keV and Compton reflection, as well as a soft blackbody-like component below 5 keV. The dominating powerlaw-like component shows a clear cutoff above 50 keV, and the spectral shape is consistent with being produced  by  thermal Comptonization in plasma of electron temperature $\sim$40 keV. At the peak of the outburst, we also detect a soft excess below 3--4 keV, which we associate with the presence of the accretion disc.

\item
The amplitude of Compton reflection is correlated with the flux. In spite of the fact, that the actual value for this amplitude is model dependent, the trend is not. This behaviour is generally consistent with the scenario where the disc recedes as the accretion rate drops and the magnetospheric radius grows. The observed sharp increase of the reflection at the highest fluxes is inconsistent with the simple circular model of the hard X-ray emitting region, but more  consistent with a narrow ring. The high reflection fraction and its sharp dependence on flux might also imply that the inner disc is puffed up because of the interaction with the magnetosphere. In any case, the observations require the inner disc radius  to be close to the stellar surface at the peak of the outburst.

\item
The spectral analysis of the phase-averaged spectra also indicates a strong correlation of the apparent emitting area with the flux
during the outburst. This result is largely  model independent.  The thermal Comptonization model with equal areas of the blackbody and Comptonization components give the area dropping from about 100 to 40 km$^2$ in the course of the outburst. The actual value for the apparent area of the emitting spot depends on the assumed ratio of the two components, and has a factor of two uncertainty.

\item
We study the variations of the pulse profiles during the outburst and their dependence of photon energy. The profile is very stable during the slow decay stage and is similar to those observed during other outbursts at similar flux level. However, we observe significant variations of the profiles at  both high and low fluxes.

At the peak stage, the deviations are stronger at higher energies. We estimate the optical depth through the accretion stream and show that at highest fluxes, it is marginally optically thick and can absorb part of the radiation at certain pulsar phase resulting in the observed dips close to the pulse maximum. The energy dependence of the effect seems to indicate that the Comptonizing region occupies a smaller area compared to the blackbody emitting spot.

When the flux drops below a certain level, after 2002 October 29,  a secondary maximum appears in the pulse profiles. This is accompanied by the jump in the phase of the fundamental.  We interpret these facts in terms of appearance of a secondary antipodal spot.  We further develop a theoretical model of the pulse profile with the varying inner disc radius. We show that varying obscuration of the antipodal spot can produce  phase jump and pulse profile variability similar to those observed.

\item
Our energy-resolved pulse profile analysis show that the pulse shapes at various energies are considerably different.  One of the quantitative measure of this effect are the time lags, which show a complicated variability during the outburst, mostly as a result of changing the pulse profile. Relative to the 2--3 keV range, the lags are negative (i.e. high-energy emission is coming earlier than the softer emission), strongly increasing up to 7--10 keV and saturating at higher energies (see Fig. \ref{fig:lag}).

The phase-resolved spectra are well described by two major components (blackbody and Comptonized tail) with varying normalizations, representing the apparent areas of the emission components. In all outburst phases, the blackbody component are lagging the Comptonized one resulting in soft time lags and explaining their energy dependence. These facts could be explained by the model where the blackbody and Comptonization emissivity patterns are different. However, the physical nature of the very strong blackbody variability observed during the FT stage remains unknown.

\item
The flux at the moment of significant change in the pulse profile, if associated with the appearance of the antipodal spot, can be used to put a constraint on the inner disc radius at that moment (see Table \ref{tab:rin}). It strongly depend on the unknown inclination  $i$, giving radii from about 20 to 35 km, when $i$ varies between 50\degr and 70\degr. The remaining uncertainty,  because of the unknown  stellar compactness and the size of the emitting spot, is about 20 per cent.

\item
The information about the flux and the inner disc radius at the moment of appearance of the antipodal spot  allows us to  estimate the stellar magnetic dipole moment $\mu\approx (9\pm5)\times10^{25}\ka^{-7/4}$ G cm$^3$, where $\ka$ is the ratio of the inner disc radius to the Alfv\'en radius.  The uncertainties here are dominated by the unknown inclination.  A similar value $\mu\approx (6\pm2)\times10^{25}\ka^{-7/4}$ G cm$^3$ is obtained from the fact that at the peak of the outburst,  the inner disc radius is in close vicinity of the neutron star as implied by the observed high reflection fraction. The corresponding surface magnetic field (for NS radii 10--15 km) is $B_0\approx (0.8\pm0.5) 10^8\ka^{-7/4}$ G. These results are    in excellent agreement with the values obtained recently from the pulsar spin-down rate (H08) assuming $\ka\sim1$. To be consistent with the results of H08, one requires  $\ka$ to deviate from unity by not more than factor of 2, which give interesting constraints on the physics of accretion onto a magnetized star.

\end{enumerate}

The analysis of the 2002 outburst of \sax\ led us to the following coherent picture of the outburst. At highest observed luminosities, the accretion disc is very close to the star. It fully blocks the view of the antipodal spot and results in a high reflection fraction. The accretion stream at this stage is marginally optically thick and absorbs part of the radiation at a phase close to the pulse maximum. The behaviour of the reflection seems to indicate a rather narrow ring-like emission region and/or puffed up inner disc. As the flux drops, during the slow decay stage, the decreasing reflection indicates that the disc is retreating. The pulse profile is however very stable, indicating that the accretion stream is now optically thin and the antipodal spot is still blocked from the view by the disc. During the rapid decay  phase, the disc is far enough and  the antipodal spot opens to the observer's view. This is  accompanied by the fast changes in the pulse profile and sharp jump in the phase of the fundamental. The outlined scenario is consistent with all the data on \sax\ obtained during its five outbursts.

There are still a number of uncertainties and unanswered questions. In our analysis and interpretation of the data, we have made somewhat contradicting assumptions. For spectral modelling, we have assumed a fixed inner disc radius of 10$\rg$, while the pulse variability and changes in the reflection indicate a varying radius. This is not a problem, as the obtained reflection fraction is not affected much by the assumption on the disc radius. For estimations of the inner disc radius and stellar magnetic moment from the flux at the moment of appearance of the antipodal spot, we assumed a circular emitting spot. On the other hand, the reflection-flux dependence indicates that the correct hotspot geometry might be closer to a ring. This difference does not affect the conclusions as long as the ring thickness is not much smaller than half of its radius, which is consistent with the numerical simulations \citep{RU04,KR05}. Our theoretical calculations of the pulse profile variations as a function of the inner radius would not be affected much in that case.

There are other effects not yet accounted in modelling. For example, we have assumed equal areas of the blackbody and Comptonization components. A dip in the high-energy profile at the peak of the outburst might indicate, however, that the high-energy component is produced in a smaller area. This discrepancy does not affect our conclusions regarding the decreasing spot area during the outburst and the estimations of the inner disc radius as we accounted for a factor of two uncertainty. However, the calculations of the pulse profile will be affected, because the retreating disc  would affect the pulse profiles at low and high energies in a different way.
This is an interesting problem for further studies. In order to reduce the uncertainties still present in the model and to get better constraints on geometrical parameters (e.g. inner disc radius), it would be useful to know the exact spot shape as a function of the magnetic inclination and the inner disc radius. A progress in this direction is possible with the detailed MHD simulations of the accretion disc-magnetosphere interaction \citep[see e.g.][]{RU04,KR05}. A further progress would also require a detailed model of the shocked region, where the high-energy radiation is produced, together with the radiative transfer modelling in order to predict the angular emission pattern from first principles.

\section*{Acknowledgments}
We are grateful to Jake Hartman for the timing solution used in the paper and useful comments.
We thank  Deepto Chakrabarty and Rudy Wijnands for helpful discussions, and the anonymous referee for the constructive criticism.
AI was supported by the Finnish Graduate School in Astronomy and Space Physics, the V\"ais\"al\"a Foundation and the Russian Presidential program for support of leading science schools (grant NSh 4224.2008.2). JP acknowledges support from the Academy of Finland grants 110792 and 127512. We also acknowledge the support of the International Space Science Institute (Bern, Switzerland), where part of this investigation was carried out.


\appendix

\section{Flux from a ring-shaped spot}
\label{app:flux}

For approximate calculation of the flux and reflection, we neglect magnetic inclination, and consider the emitting region in the form of a ring symmetric around the stellar rotational axis. The boundaries  of the ring are described by colatitudes $\rho_{\min}$ and  $\rho_{\max}$. We assume that the light from the antipodal ring below the equator is blocked by the accretion disc.

Let us first compute the observed flux from the ring.  Consider a surface element $\rmd S=\rstar^2 \rmd \cos\theta \rmd \phi$ at a slowly rotating star. We neglect here the effects related to rapid rotation because the flux averaged over the ring is affected by them very little. The element position is described by the unit vector $\bmath{n}=(\sin \theta\cos\phi, \sin \theta\sin\phi,\cos\theta)$ that points to it from the star center. Let the unit vector along the line of sight be $\bmath{k}=(\sin i, 0, \cos i)$ with $i$ being the inclination.  The angle between $\bmath{n}$ and the line of sight is denoted by $\psi$ so that
\be \label{eq:cospsi}
\cos\psi=\bmath{k}\cdot \bmath{n} = \cos i\ \cos\theta+\sin i\ \sin \theta\ \cos\phi.
\ee
The observed flux from this surface element  in approximation of \citet{b02} for light bending is (see PG03; \citealt{PB06}):
\be \label{eq:fluxele}
\rmd F(i,\theta,\phi)=  (1-u)^2  \frac{\rmd S}{D^2}  I_*(\alpha) \cos\alpha ,
\ee
where the angular distribution of intensity at the stellar surface $I_*(\alpha)$ is assumed to follow the dependence (\ref{eq:intani}),
$\alpha$ is the angle of the photon emission relative to the stellar normal related to other parameters as
\be \label{eq:alphapsi}
\cos\alpha \approx u + (1-u) \cos\psi = Q+ U \cos\phi ,
\ee
with $Q$ and $U$ given by equation (\ref{eq:U}).

Integrating (\ref{eq:fluxele}) over the ring surface, we get
\be \label{eq:flux_ring}
F(i) = I_0 \frac{\Sigma}{D^2}  (1-u)^2     \zeta
\ee
where   $\Sigma=\rstar^2 2 \pi (\mu_2-\mu_1)$ is the ring area, $\mu_{1}=\cos\rho_{\max}$, $\mu_{2}=\cos\rho_{\min}$, and
\beq \label{eq:bracket}
\zeta &=&
u+ \frac{1-u}{2}  (\mu_1+\mu_2) \cos i\ \nonumber \\
& +& \ani \Big[ u^2 + u (1-u) (\mu_1 + \mu_2)\cos i\  + \frac{1}{2}(1-u)^2\sin^2i \Big]  \nonumber \\
&+& \ani \frac{1}{6}(1-u)^2 (3\cos^2i-1) \ (\mu_1^2 + \mu_1 \mu_2 + \mu_2^2)  ,
 \eeq
which is strictly valid when the whole ring is visible.

The total emitted luminosity at the stellar surface (for two symmetric about equator, ring-shaped emitting regions) is
\be
L_{*} = 2\Sigma\ 2\pi \int_0^1 I_*(\alpha) \cos\alpha\ \rmd\cos\alpha=
2\Sigma\ I_0 \ 2\pi \left( \frac{1}{2}+\frac{\ani}{3}\right),
\ee
and the total luminosity at infinity $L=(1-u)L_{*}$. We thus get the anisotropy correction
factor relating the observed flux to the luminosity in equation (\ref{eq:LFfang}):
\be \label{eq:fangcomp}
 f_{\rm ang} = \frac{L}{4\pi D^2 F(i)} = \frac{ \frac{1}{2}+\frac{\ani}{3}}{(1-u)\zeta} .
\ee

\begin{figure}
\begin{center}
\centerline{\epsfig{file=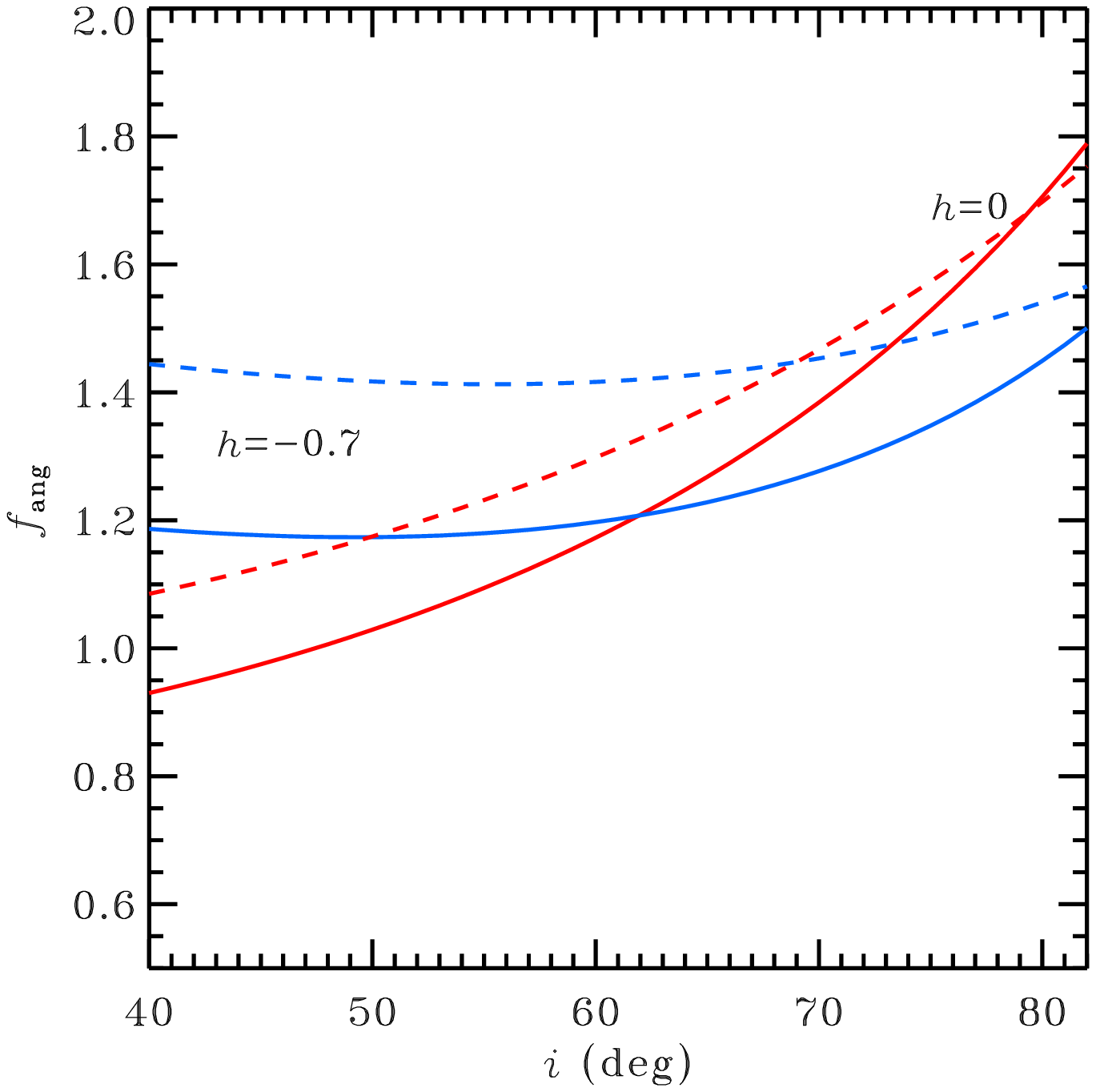,width=7.2cm}}
\end{center}
\caption{The anisotropy correction factor as a function of inclination computed from equations (\ref{eq:bracket}) and (\ref{eq:fangcomp}).
The solid curves correspond to stellar parameters  $M=1.4\msun$  and $\rstar=12$ km, while
the dashed curves to $M=1.7\msun$  and $\rstar=11$ km. More inclined curves are for blackbody-like radiation
pattern with $\ani=0$ and almost horizontal  curves are for $\ani=-0.7$.
Here a circular spot centred at the rotational pole with $\rapp=5$ km is assumed.
}
\label{f:fang}
\end{figure}

  As an example, let us consider a circular spot (i.e. $\rho_{\min}=0$) centred at the rotational pole. For a given apparent spot size $\rapp$, we compute the physical spot angular size $\rho_{\max}=\rho$ from equation (\ref{eq:sinrho}). The dependence of $f_{\rm ang}$ on the inclination is shown in Fig. \ref{f:fang}.  Variations of $\rapp$ within 3--7 km, lead to only 2--3 per cent changes in $f_{\rm ang}$.

\section{Reflection from the accretion disc}
\label{app:refl}

Let us now estimate the reflection amplitude expected for the emitting ring. By definition it is given by the ratio of the luminosity reflected from the disc to that directly observed:
\be \label{eq:refl_flux}
\refl = \frac{L_{\rm refl}}{4\pi D^2 F(i)} = f_{\rm ang} P_{\rm refl},
\ee
where $P_{\rm refl}$ is the fraction of emitted photons that hit the disc surface at radii $r>\rin$.

Because of the axial symmetry, the surface element in a ring is described only by colatitude with the corresponding unit vector pointing towards it, $\bmath{n}=(\sin \theta, 0,\cos\theta)$. The photon direction (in the frame related to the element, with $z$-axis along its normal) is described by two angles, polar angle $\alpha$ and azimuth $\varphi$   measured from the southern direction of the meridian. 
At large distances from the star the photon direction is given by vector $\bmath{k}$. For a given $\alpha$, the angle $\psi$ its makes to $\bmath{n}$ can be obtained from equation (\ref{eq:alphapsi}).  Using spherical trigonometry it is easy to show that
\be
\bmath{k}=\left( \begin{array}{c}
\cos\psi  \sin\theta +\sin \psi \cos\theta \cos\varphi \\
\sin \psi\sin\varphi  \\
\cos \psi \cos\theta -\sin \psi  \sin\theta  \cos\varphi
\end{array}
\right) .
\ee
The necessary condition that the trajectory crosses the disc is 
\be
\cos \psi \cos\theta < \sin \psi  \sin\theta  \cos\varphi ,
\ee
that translates to the limits on $\alpha$ and $\varphi$. In the stellar polar region $\sin\theta<\kappa\equiv  u/(1-u)$
(existing if $u<1/2$), the limits are
\be \label{eq:alphi1}
\begin{array}{ll}
 0<\varphi<2\pi, & \quad\mbox{if}\quad 0<\cos\alpha<\cos\alpha_- ,   \\
\cos\varphi>\cos\varphi_0, & \quad\mbox{if} \quad \cos\alpha_-<\cos\alpha<\cos\alpha_+ ,
\end{array}
\ee
where $\cos\alpha_{\pm}=u\pm(1-u)\sin\theta$ and $\cos\varphi_0=\cot \psi\cot\theta$. Thus photons emitted close to the surface will always cross the disc plane irrespectively of the azimuth, and those emitted close to the zenith $\cos\alpha>\cos\alpha_+$ do not cross it at all. In the intermediate case, photons have to be emitted in the southward direction to cross the disc plane. Outside the polar region $\sin\theta>\kappa$, the limits are
\be \label{eq:alphi2}
\cos\varphi>\cos\varphi_0,  \quad\mbox{if} \quad 0<\cos\alpha<\cos\alpha_+ .
\ee

Now let us compute the radius where  the photon trajectory crosses the disc plane. The unit vector
along the intersection line of the disc and photon trajectory planes is
\be
\bmath{d} = \frac{(\cos\varphi,\ \cos\theta\sin\varphi ,\ 0)}{\sqrt{1-\sin^2\theta \sin^2\varphi} } ,
\ee
and thus the angle it makes  with $\bmath{k}$ is given by
\be \label{eq:cospsidisc}
\cos\psi_{\rm d} = \bmath{d} \cdot \bmath{k} =
\frac{\sin\psi\cos\theta+\cos\psi\sin\theta\cos\varphi}{\sqrt{1-\sin^2\theta \sin^2\varphi} }   .
\ee
For the given emission angle $\alpha$ and the  impact parameter
\be \label{eq:impact}
b=\rstar\sin\alpha/\sqrt{1-u},
\ee
an approximate photon trajectory is \citep{b02}:
\be \label{eq:rpsi}
r(\psi) =\Big[\frac{\rg^2(1-\cos\psi)^2}{4 (1+\cos\psi)^2}+\frac{b^2}{\sin^2\psi}\Big]^{1/2}-
\frac{\rg(1-\cos\psi)}{2(1+\cos\psi)} .
\ee
Substituting here $\psi=\psi_{\rm d}$, we obtain the disc radius, where photon trajectory crosses the disc  plane.

Calculation of the reflected fraction now involves simple integration over the ring area (upper hemisphere) and  photon emission angles:
\be
P_{\rm refl}\!\!=\!\! \frac{4\pi\rstar^2}{L_{*}} \!\!  \!\int\! \rmd \cos\theta\!\!
 \int \!\!  I_*(\alpha)\cos\alpha\ \rmd\cos\alpha\! \int \!\! \rmd \varphi \ H (r[\psi_{\rm d}]-\rin) ,
\ee
where $H$ is the Heaviside step function and the limits on $\alpha$ and $\phi$ are given by conditions (\ref{eq:alphi1}) and (\ref{eq:alphi2}).

\section{Antipodal spot eclipse by the accretion disc}
\label{app:discabs}

The pulse profiles produced by the spots on a rapidly rotating star are computed following techniques described in PG03 and \citet{PB06}. However, the effect of the accretion disc needs to be accounted for, because the photons emitted by the lower antipodal spot can be absorbed on the way to the observer by the disc. We divide the spot into a number of elements and compute the light curve from each of it independently.  The position of each element can be described, as in Appendix \ref{app:flux},  by the unit vector $\bmath{n}=(\sin \theta\cos\phi, \sin \theta\sin\phi,\cos\theta)$, where now $\theta>\pi/2$. The angle $\psi$ this vector makes with the direction to the observer $\bmath{k}=(\sin i, 0, \cos i)$  is given by equation (\ref{eq:cospsi}).

The element in principle (without the disc) is visible to the observer at all phases if $\cos(\theta+i)>-\kappa$, not visible at all if $\cos(\theta-i)<-\kappa$, and visible only when $\cos\phi> - Q/U$ 
if $\cos(\theta+i)<-\kappa<\cos(\theta-i)$ \citep[see][ for details]{PB06}. If the element with the given $(\theta,\phi)$ can be visible, we substitute $\psi$ given by equation (\ref{eq:cospsi}) to the light bending formula (\ref{eq:alphapsi}) getting the emission angle $\alpha$ and the impact parameter $b$ using equation (\ref{eq:impact}), which fully define the photon trajectory. Its plane intersects with the disc plane along the line described by the unit vector
\be
\bmath{d} = \frac{(-\cos\theta\sin i+ \cos i\sin\theta\cos\phi, \ \sin\theta\cos i \sin\phi,  \ 0)}
{\sqrt{\cos^2i+\cos^2\theta-2\cos i \cos \theta\cos\psi} } .
\ee
The angle it makes with $\bmath{k}$ is
\be \label{eq:cospsid_anti}
\cos\psi_\mathrm{d}=  \bmath{d} \cdot \bmath{k} =
\frac{\cos i\cos\psi-\cos\theta}{\sqrt{\cos^2 i+\cos^2\theta-2\cos i\cos\theta\cos\psi}}.
\ee
We find the radius, where trajectory crosses the disc $r(\psi_\mathrm{d})$, using equation (\ref{eq:rpsi}). If it is larger than $\rin$, the element is invisible.

\label{lastpage}

\end{document}